\documentclass[a4paper,fleqn]{cas-dc}

\usepackage[style=numeric,sorting=none,url=false,backend=biber]{biblatex}
\hypersetup{
    pdftitle={Impact of Structural Design on Magneto-Mechanical Resonators: The Case of a Jewel Bearing Variant},
    }

\newcommand{\plotspace}{\vspace{-20pt}}

\usepackage{orcidlink}
\newcommand{\orcidmark}[1]{\kern-0.15em\textsuperscript{\orcidlink{#1}}}

\ExplSyntaxOn
\RenewDocumentCommand \printorcid { }
{
  \seq_if_empty:NF \g_stm_orcid_seq
  {
    \group_begin:
      \tex_let:D \thefootnote \relax \footnotetext
      { \raggedright \textsc{orcid}(s):\c_space_token
        \seq_use:Nn \g_stm_orcid_seq { ;~ } }
    \group_end:
  }
}
\ExplSyntaxOff

\begin{document}
\let\WriteBookmarks\relax
\def\floatpagepagefraction{1}
\def\textpagefraction{.001}

\title[mode = title]{Impact of Structural Design on Magneto-Mechanical Resonators: The Case of a Jewel Bearing Variant}

\shorttitle{Impact of structural design on magneto-mechanical resonators}

\shortauthors{D. Conrad et~al.}

\author[1,2]{David Conrad}[suffix={\orcidmark{0009-0009-2156-9419}}]
\cormark[1]
\ead{david.conrad@uni-rostock.de}
\credit{Conceptualization, Methodology, Data curation, Resources, Software, Formal analysis, Investigation, Validation, Visualization, Writing -- original draft}

\affiliation[1]{organization={Metrology Lab, University of Rostock},
            city={Rostock},
            country={Germany}}

\author[1,2]{Jan-Philipp Scheel}[suffix={\orcidmark{0000-0002-7840-4879}}]
\credit{Conceptualization, Methodology, Resources, Writing -- review \& editing}

\affiliation[2]{organization={Fraunhofer Research Institution for Individualized Medical Technology and Engineering IMTE},
            city={Lübeck},
            country={Germany}}

\author[1]{Frauke H. Niebel}[suffix={\orcidmark{0000-0003-3394-0122}}]
\credit{Conceptualization, Methodology, Software, Writing -- review \& editing}

\author[1]{Hossein Shabanalinezhad}[suffix={\orcidmark{0009-0007-2693-8899}}]
\credit{Conceptualization, Methodology, Writing -- review \& editing}

\author[2]{Nora Timm}[suffix={\orcidmark{0009-0001-2591-2925}}]
\credit{Conceptualization, Writing -- review \& editing}

\author[3]{Sarah Reiss}[suffix={\orcidmark{0009-0006-4015-1200}}]
\credit{Software, Writing -- review \& editing}

\affiliation[3]{organization={Institute for Biomedical Imaging, Hamburg University of Technology},
            city={Hamburg},
            country={Germany}}

\author[3]{Fabian Mohn}[suffix={\orcidmark{0000-0002-9151-9929}}]
\credit{Resources, Writing -- review \& editing}

\author[2,3]{Tobias Knopp}[suffix={\orcidmark{0000-0002-1589-8517}}]
\credit{Software, Writing -- review \& editing}

\author[1]{Matthias Graeser}[suffix={\orcidmark{0000-0003-1472-5988}}]
\credit{Conceptualization, Methodology, Resources, Supervision, Project administration, Funding acquisition, Writing -- review \& editing}

\cortext[1]{Corresponding author}

\nonumnote{Preprint submitted 30 July 2026, accepted 13 October 2026 for publication in \textit{Measurement: Sensors} (Elsevier)}

\nonumnote{This work was supported by a scholarship from the Landesgraduiertenförderung Mecklenburg-Vorpommern (state graduate funding program of Meck\-len\-burg-Vorpommern, Germany).}

\begin{abstract}
Magneto-mechanical resonators (MMRs) are passive, wirelessly read sensors whose small size, miniaturizability, and low cost make them attractive for tracking and for sensing physical and chemical quantities. Their operation relies on a permanent-magnet rotor whose mechanical resonance encodes the sensing and tracking information. The bearing suspending this rotor is therefore decisive both for in-operation performance and for manufacturability. The original design suspends the rotor on a thin thread, which is nontrivial to assemble. This work investigates the impact of that structural design choice by introducing an alternative bearing in which the spherical rotor magnet rests in a cup-shaped industrial jewel. A dynamic model of this Jewel-MMR is derived, including the angle-dependent dry friction torque at the jewel contact. From it, a geometric trade-off between friction torque and resilience against unwanted oscillation modes is identified. Since the conventional quality factor loses its meaning under dominant dry friction, a two-tiered framework is proposed that compares MMR variants by estimation precision at equal magnet size and natural frequency. Both variants are characterized at a fixed pose on a Helmholtz-coil detection platform. The Jewel-MMR is self-aligning, requires fewer adhesive bonds and alignment steps, and the prototypes built here were assembled in less than half the time. Its response signal decays faster, so that under laboratory noise conditions the natural frequency and the orientation are recovered more precisely from the thread variant; these orientation figures quantify repeatability, not absolute accuracy. The jewel bearing, however, tolerates larger deflection angles and thus a stronger signal, which reverses this relation above a noise level determined in a simulation-based robustness analysis, indicating an advantage at reduced signal-to-noise ratio.
\end{abstract}

\begin{graphicalabstract}
\includegraphics[width=\textwidth]{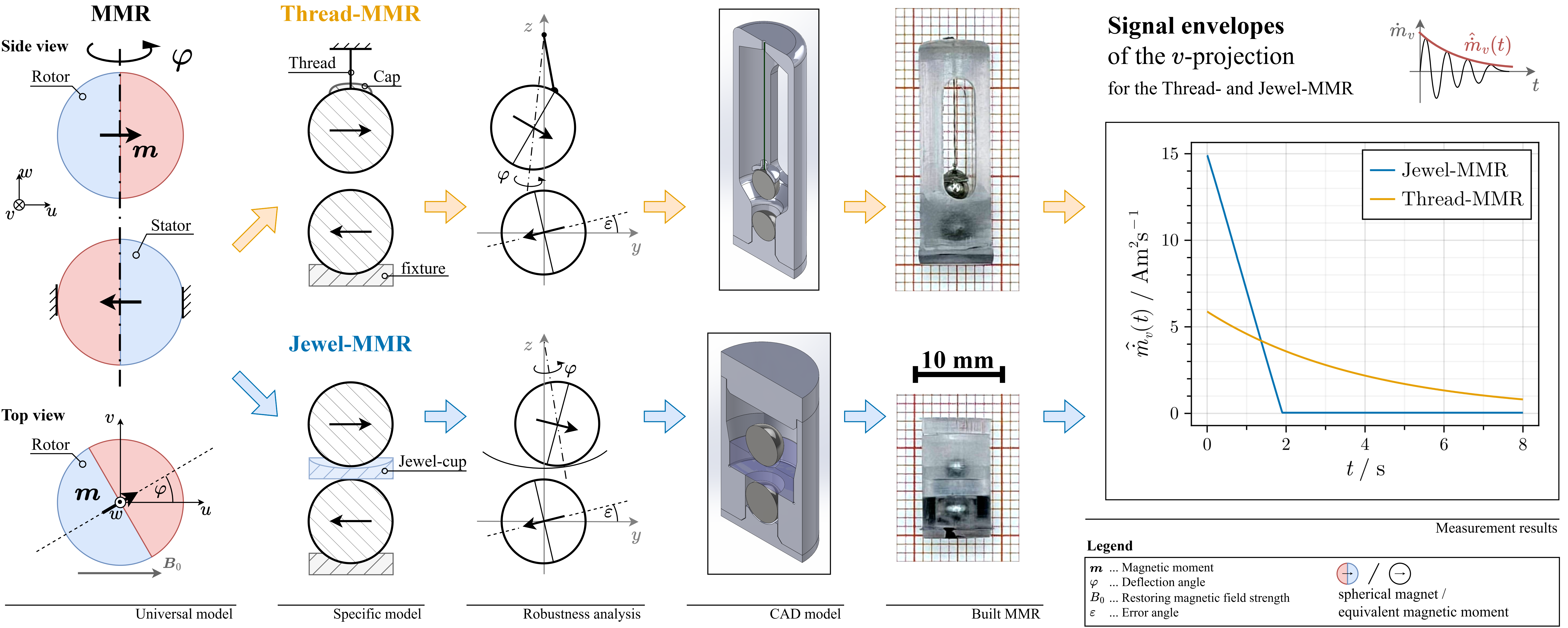}
\end{graphicalabstract}

\begin{highlights}
\item Jewel bearing introduced as self-aligning alternative to the MMR thread suspension
\item Dynamic model derived with angle-dependent dry friction at the jewel contact
\item Geometric trade-off between friction torque and predicted side mode susceptibility
\item Two-tiered framework compares MMR variants by sensing and tracking precision
\item Jewel-MMR simplifies manufacturing and gains a precision advantage at elevated noise
\end{highlights}


\maketitle

\section{Introduction}
\label{sec:introduction}

Magneto-mechanical resonators (MMRs) are passive, battery-free sensors that are read out wirelessly through their magnetic fields. Because they require neither an internal power source nor electrical connections, they can be read out remotely through tissue and other non-conductive materials. In addition, they can, in principle, be miniaturized to sub-millimeter scales at low cost. Together, these properties address several limitations of established tracking and sensing technologies \cite{gleich_miniature_2023}. In minimally invasive surgery, for example, the position and orientation of instruments such as catheters, guidewires, or biopsy needles must be known inside the body. Here, optical line-of-sight tracking is unavailable, cabled electromagnetic trackers may not be able to reach every location, and imaging-based approaches are costly or rely on ionizing radiation. Wireless radiofrequency markers, another wireless tracking technique, require a minimum size of roughly a centimeter to accommodate a communication antenna \cite{gleich_miniature_2023}. Beyond localization, MMRs can wirelessly report physical quantities such as temperature, pressure, or the presence of specific chemicals. This makes them suited for both medical in-body physiological monitoring and industrial use cases. The latter involves tracking and condition monitoring of goods, where an MMR can act as a miniature, electronics-free counterpart to an RFID tag \cite{gleich_miniature_2023} or pressure sensing in process engineering \cite{merbach_wireless_2025}. \citeauthor{jing_frequency_2023} also demonstrated that a similar concept can be used to generate ultralow-frequency magnetic signals for through-ground or through sea-water data transmission \cite{jing_frequency_2023}.

At its core, the MMR sensing technology consists of the passive MMR itself and a detection platform. The latter first drives the MMR via alternating magnetic fields into an excited state during a dedicated excitation window. Subsequently, during the receive window, it picks up the magnetic response signal. It originates from a mechanically oscillating permanent magnet inside the resonator, whose restoring torque usually relies on one or more additional magnets \cite{gleich_miniature_2023,kanj_design_2022}. Other publications also use mechanical \cite{fischer_miniature_2024} or hybrid \cite{li_pivot_2025} systems, but these are not further discussed in detail here. The spatial distribution of the receive signal's spectral components across the detection coils encodes the position and full orientation of the resonator, which can be exploited for tracking. For sensing applications, any physical quantity that modulates the distance between the involved magnets shifts the sensor's natural frequency, which can be sensed remotely. In practice, effects such as thermal expansion for temperature, a compressible housing for pressure, or a responsive material for radiation or chemical markers are used. This dual capability of simultaneous tracking and sensing from a single passive element is the central appeal of the MMR technology. \cite{gleich_miniature_2023}

The concept of such a magnet-based resonator was introduced by \citeauthor{gleich_miniature_2023}~\cite{gleich_miniature_2023}, and several groups have investigated variants of the underlying bearing and magnet geometry \cite{faltinath_natural_2025, knopp_empirical_2024, li_pivot_2025, kanj_design_2022}. Common to all of these designs is a rotor magnet that must retain a single rotational degree of freedom about the main axis while all remaining degrees of freedom are suppressed. The bearing realizing this constraint is thus the central design element. It governs both the achievable resonance quality and the effort required to manufacture the device. The original solution suspends the rotor on a thin thread \cite{gleich_miniature_2023, faltinath_natural_2025}. While this approach achieves high quality factors, it entails several practical drawbacks: The thread must be attached and aligned with high precision. Also, the use of curing adhesive that seeps into the thread reduces its effective length and therefore limits how far the assembly can be miniaturized. But most critically for long-term robustness, a torn thread irreversibly destroys the sensor function. Outside the MMR context, bearings for oscillating and rotating precision systems are already a well-studied field. Of particular interest here is the jewel bearing, in which a hard tip rests on the polished surface of a sapphire or ruby jewel, so that the load is transmitted over a contact area small enough to keep the friction torque low. Two operating regimes are established for it. In high-speed rotating machinery such as flywheels, spindles, and centrifuges, pivot jewels rotate continuously under oil-bath lubrication and are usually relieved by an additional magnetic bearing that carries part of the rotor weight \cite{dai_friction_2013}. In instruments whose moving element oscillates about a nearly stationary contact point, such as galvanometers, compasses, indicating meters, and mechanical watches, the same bearing type is lightly loaded and operated dry \cite{krause_lager_2018, de_la_presilla_oscillating_2023}. Since MMRs share a similar oscillatory behavior, this work addresses only the latter regime. 

A second aspect the MMR shares with related systems concerns how such a resonator is characterized. Both the magnetic restoring torque and, for the variant proposed here, the friction at the bearing contact make the MMR a nonlinear oscillator, so that a single linear parameter cannot describe it. For passive elasto-magnetic suspensions, \citeauthor{bonisoli_identification_2007}~\cite{bonisoli_identification_2007} compared a non-parametric restoring force mapping method with a direct parametric estimation technique and found both to agree well with experimental data, the non-parametric one somewhat more accurately and without requiring a model to be assumed beforehand. It is, however, applicable only where the instantaneous displacement and velocity of the oscillator is available, which for the MMR would require additional measurements hardware. In this work, the characterization therefore rests on the time-domain response signal alone, following \citeauthor{reiss_parameter_2026}~\cite{reiss_parameter_2026} and \citeauthor{kanj_design_2022}~\cite{kanj_design_2022}.

Given the current state of the art, the present work investigates how the structural design of the bearing impacts the performance and manufacturability of MMRs, pursuing two goals. First, it introduces a dry operated jewel bearing as an alternative to the thread suspension, in which the spherical rotor magnet rests in a cup-shaped industrial jewel. The aim is explicitly not to claim a universally superior design, but to present a complementary option and to expose the trade-offs and pitfalls inherent to each bearing concept. Second, and more generally, it addresses the question of how different MMR variants can be compared in a fair and meaningful way. It is argued that such a comparison must span the entire life-cycle of the sensor, most notably the manufacturing process, which strongly influences cost, achievable tolerances, and ultimately the resonator quality, rather than focusing on in-operation metrics alone.

Using the Jewel-MMR as a representative example for alternative bearing concepts, the contributions of this work are: (i) a dynamic model of the Jewel-MMR in the time-domain including the angle-dependent dry friction torque at the jewel contact; (ii) a discussion of suitable quality measures for MMRs and a two-tiered evaluation framework that combines intra-variant damping measures with a fair, but size-dependent inter-variant comparison based on the estimation precision of these quantities; (iii) an analysis of how manufacturing defects translate into susceptibility to unwanted side mode excitation, revealing a geometric design trade-off for the jewel cup; and (iv) an experimental comparison of a Jewel- and a Thread-MMR of equal size with respect to manufacturability and to the precision with which the natural frequency and the orientation, that is the quantities underlying sensing and tracking, are estimated at a fixed pose and under varying noise levels. It should be emphasized that no complete sensing or tracking task is carried out in this work. Neither is a measurand applied to the resonator, nor is its position estimated. Instead, both variants are measured repeatedly at a fixed pose, and the two quantities from which an application would derive its sensing and tracking information are compared with respect to their estimation precision.

\section{Methods}

\subsection{Overview of MMR variants}

\begin{figure}[pos=t]
\centering
\includegraphics[width=.85\columnwidth]{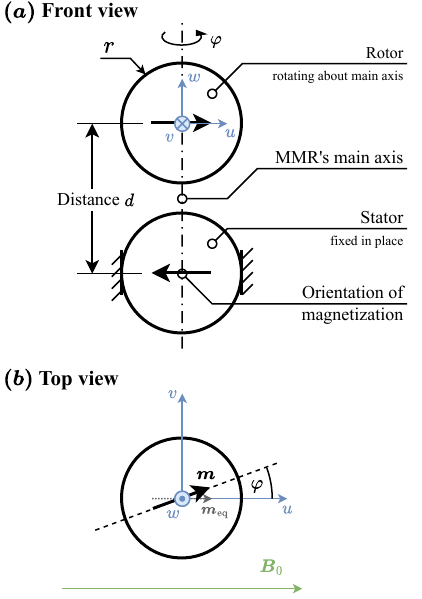}
\caption{Definitions regarding the standard MMR principle with two spherical magnets of radius $r$ in \textit{(a)} a frontal view perpendicular to the MMR's main axis and \textit{(b)} a view from above, visualizing the rotor magnet with its magnetic moment~$\boldsymbol{m}$ at deflection angle $\varphi$ within the local $uvw$ coordinate system. $\boldsymbol{B}_0$ is the stator's magnetic field at the rotor's position and $\boldsymbol{m}_\mathrm{eq}$ marks the equilibrium orientation of the rotor.}
\label{fig:mmr-definitions}
\end{figure}

At the conceptual level, MMRs that solely rely on a magnetic restoring torque usually consist of two permanent magnets \cite{gleich_miniature_2023}. \citeauthor{faltinath_natural_2025} used cylindrical magnets with diametrical magnetization as they maximize the magnetic moment with the same overall height \cite{faltinath_natural_2025}. During assembly, the flat surfaces of a cylinder can be advantageous when aligning the magnets. Regardless, this work utilizes spherical magnets, for which the magnetic forces, torques, and energies can be accurately modeled with magnetic dipoles $\boldsymbol{m} = m_\mathrm{r} \, \boldsymbol{e}_m$ \cite{edwards_interactions_2017}. To assist the alignment process of the magnets in this work, interface caps and fixtures are used, further described in Section \ref{sec:experiments}.

As shown in Fig.~\ref{fig:mmr-definitions}\textit{a}, the magnetic spheres with radius $r$ or their equivalent dipoles are stacked on top of each other along the MMR's main axis. The magnetization orientation of the magnets is diametrical with respect to this axis. One magnet is fixed with regard to the MMR's housing and is referred to as stator, while the other, called rotor, can move. An ideal bearing constrains this motion to a single remaining degree of freedom: lossless rotation about the main axis. In operation, the rotor typically does not complete full rotations, but oscillates around its equilibrium position. At this point of equilibrium of the magnetic forces, the magnetic dipoles face opposite directions. As shown in Fig.~\ref{fig:mmr-definitions}\textit{b}, the rotor's deflection from its equilibrium position is described by the angle $\varphi$ in the sensor's local $uvw$-coordinate system. The mathematical description for the restoring magnetic torque $\boldsymbol{T}_\mathrm{Magn.}$ acting on the rotor dipole moment, expressed as
\begin{equation}
    \boldsymbol{m}_{uvw}(\varphi) = m_\mathrm{r} \ (\cos \varphi \ \boldsymbol{e}_u + \sin \varphi \ \boldsymbol{e}_v)
    \label{eq:m_uvw(phi)}
\end{equation}
deflected by $\varphi$ about the $w$-axis is given by
\begin{align}
    \label{eq:T_m}
    \boldsymbol{T}_\mathrm{Magn.} &= \boldsymbol{m}_{uvw}\times\boldsymbol{B}_0
        = \begin{pmatrix} m_\mathrm{r} \cos \varphi \\ m_\mathrm{r} \sin \varphi \\ 0 \end{pmatrix} \! \times \! \begin{pmatrix} B_0 \\ 0 \\ 0 \end{pmatrix} \\
    \label{eq:T_m_2}
        &= - m_\mathrm{r} B_0 \sin\varphi \; \boldsymbol{e}_w \; ,
\end{align}
where $\boldsymbol{B}_0 =\boldsymbol{B}_\mathrm{stator}(\boldsymbol{p}_\mathrm{rotor}) = B_0 \ \boldsymbol{e}_u$ is the stator's magnetic field at the rotor's position $\boldsymbol{p}_\mathrm{rotor}$ \cite{gleich_miniature_2023}.
In practice, bearings that deviate from this ideal scenario, such as the thread bearings mentioned above, are common \cite{gleich_miniature_2023}. They technically feature oscillations other than the intended one, further referred to as side modes. However, this is generally compensated for by the targeted excitation of the MMRs' main mode only. During measurements, a homogeneous magnetic excitation field $\boldsymbol{B}_\mathrm{Exc.}(t)=B_\mathrm{Exc.,u}(t)\,\boldsymbol{e}_u+B_\mathrm{Exc.,v}(t)\,\boldsymbol{e}_v$ perpendicular to the main axis of the MMR induces a torque in the rotor that meets exactly this requirement. Analogous to eq.~\eqref{eq:T_m}, the resulting excitation torque $\boldsymbol{T}_\mathrm{Exc.}=\boldsymbol{m}_{uvw}\times\boldsymbol{B}_\mathrm{Exc.}(t)$ has only a $w$-component.

During the receive window of the measurement sequence and under the assumption of the previously considered ideal conditions, the unconstrained equation of motion for the MMR's response signal can be derived from the sum of all torques about the $w$-axis to
\begin{equation}
    \underbrace{I \ddot\varphi}_{T_\mathrm{Inertia}} + \underbrace{m_\mathrm{r} B_0 \sin\varphi}_{T_\mathrm{Magn.}} = 0 \; .
    \label{eq:ODE}
\end{equation}
By defining the natural angular frequency for the undamped oscillation as
\begin{equation}
    \omega_\mathrm{nat}:=\sqrt{\frac{m_\mathrm{r}B_0}{I}} \; ,
    \label{eq:omega_nat}
\end{equation}
it follows that
\begin{equation}
    \ddot\varphi + \omega_\mathrm{nat}^2 \sin\varphi = 0 \; .
    \label{equ:ODE_norm}
\end{equation}
Equation~\eqref{eq:ODE} describes an anharmonic oscillator analogous to a gravitational pendulum. For finite deflection angles, the instantaneous angular frequency $\omega$ satisfies $\omega < \omega_\mathrm{nat}$ and approaches $\omega_\mathrm{nat}$ in the limit $\varphi \to 0$. In this small-angle regime where $\sin\varphi \approx \varphi$, the expression simplifies to the governing equation of a harmonic oscillator, resulting in an undamped oscillation at the natural frequency \cite{reiss_parameter_2026}.

The detection of the MMR's oscillation relies on the measured voltage induced in the receive coils of the detection platform by the temporal change of the MMR's magnetic moment $\frac{\mathrm{d}}{\mathrm{d}t}\boldsymbol{m}(t,\,\boldsymbol{R})$. Here,
\begin{equation}
    \boldsymbol{m}(t,\,\boldsymbol{R}) := \boldsymbol{m}_{xyz}(t,\,\boldsymbol{R}) = \boldsymbol{R} \, \boldsymbol{m}_{uvw}(\varphi(t))
\end{equation}
is the magnetic moment of the rotor at time $t$ in world coordinates $xyz$ with the rotation matrix $\boldsymbol{R}$, describing the orientation of the local MMR coordinate system towards this global reference frame.

Now, let $\boldsymbol{P}(\boldsymbol{p}_\mathrm{rotor}):\mathbb{R}^3\to\mathbb{R}^{L\times3}$ be the coil sensitivity matrix at the MMR's position $\boldsymbol{p}_\mathrm{rotor}$ in world coordinates for a sensing system that consists of $L$ receive coils. Its $l$-th row holds the field per unit current $\boldsymbol{b}_l(\boldsymbol{p})^\top$ that coil~$l$ would produce at~$\boldsymbol{p}$, which by reciprocity equals the coil's receive sensitivity. Then, according to Faraday's induction law the induced voltages $\boldsymbol{u}\to\mathbb{R}^L$ are described by
\begin{equation}
    \boldsymbol{u}(t, \, \boldsymbol{p}_\mathrm{rotor}, \, \boldsymbol{R})
    = - \boldsymbol{P}(\boldsymbol{p}_\mathrm{rotor}) \, \frac{\mathrm{d}\boldsymbol{m}(t, \, \boldsymbol{R})}{\mathrm{d}t}
    \label{eq:u(t,p,R)}
\end{equation}
with
\begin{equation}
    \frac{\mathrm{d}\boldsymbol{m}(t, \, \boldsymbol{R})}{\mathrm{d}t}
    = \dot{\boldsymbol{m}}(t, \, \boldsymbol{R})
    = \boldsymbol{R}
    \begin{pmatrix}
        -m_\mathrm{r} \,\dot\varphi(t) \sin(\varphi(t)) \\
        m_\mathrm{r} \,\dot\varphi(t) \cos(\varphi(t)) \\
        0
    \end{pmatrix} \; .
    \label{eq:projection}
\end{equation}
Thus, each receive channel of the detection platform contains a superposition of signal contributions from the local $u$- and $v$-components of the oscillating magnetic moment, with the coil sensitivities $\boldsymbol{P}(\boldsymbol{p}_\mathrm{rotor})$ acting as projection weights. Following \citeauthor{reiss_parameter_2026}, the $v$-component yields an oscillating signal at the instantaneous frequency $\omega$ and its odd harmonics, while the $u$-component contributes at $2\omega$ and higher even harmonics, a consequence of the respective sine and cosine projections of the deflection angle $\varphi$ \cite{reiss_parameter_2026}.

While the ideal dynamics assume a frictionless environment, the real-world implementation of the rotor's bearing imposes an actual damping and therefore significantly influences the equation of motion \eqref{eq:ODE}.

\begin{figure}[pos=t]
    \centering
    \includegraphics[width=.85\columnwidth]{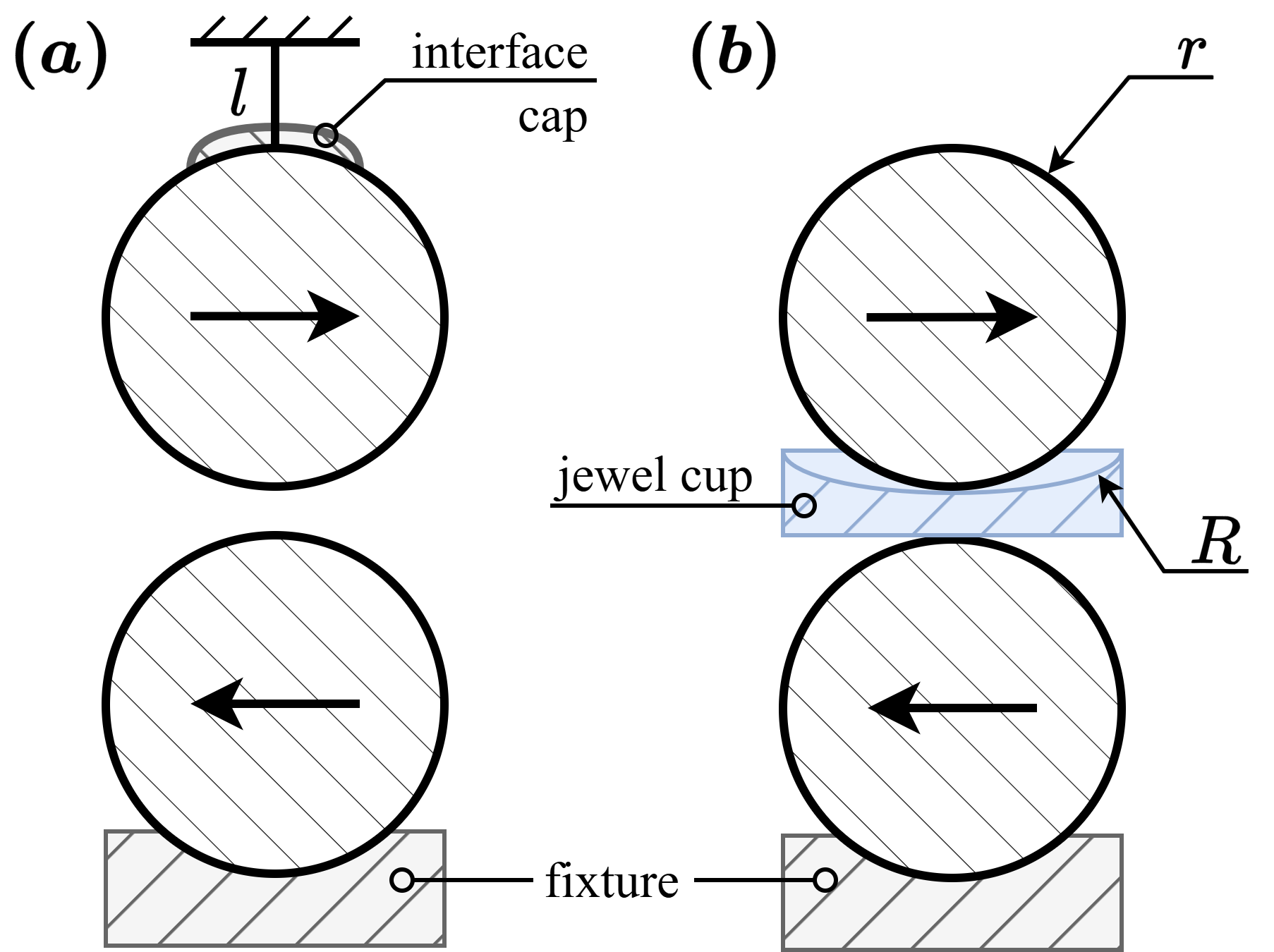}
    \caption{Schematics of MMR variants in sectional view: \textit{(a)}~Thread bearing MMR with thread length $l$ and \textit{(b)} Jewel bearing MMR with jewel-cup radius $R$. Arrows mark the magnetization orientation of the spherical magnets of radius~$r$.}
    \label{fig:mmr-variants}
\end{figure}

\subsubsection{MMRs with thread bearing}
\label{sssec:mmrs_with_thread_bearing}

Figure \ref{fig:mmr-variants}\textit{a} depicts the concept of a thread-bearing-based MMR. The stator is fixed in place as described above, marked with a gray colored cylindrical fixture that connects to the housing. The rotor is suspended by a thread of length $l$ that connects to the magnet's neutral line on one side and to the MMR's housing on the other side. Like the magnets, the thread is also aligned with the main axis, that is the axis of rotation. The thread opposing the attractive magnetic force of the stator holds the rotor magnet in its desired equilibrium position, even when deflected by the angle $\varphi$ with $\vert\varphi\vert<90^\circ$.

The equation
\begin{equation}
    I \ddot\varphi + \underbrace{c \dot\varphi}_{T_\mathrm{Viscous}} + m_\mathrm{r} B_0 \sin\varphi = 0
    \label{eq:Thread_ODE}
\end{equation}
with the viscous damping coefficient $c$ models deviations for a non-ideal, damped case of a real-world Thread-MMR by adding a term to the equation of motion, which accounts for viscous damping components. These include, among others, air friction due to laminar flow, eddy currents in the metallic coatings of the magnets, and energy dissipation in the thread \cite{gleich_miniature_2023}. Normalizing and substituting analogously to equations~\eqref{eq:ODE} and~\eqref{equ:ODE_norm} yields
\begin{equation}
    \ddot\varphi + \frac{\omega_\mathrm{nat}}{Q} \dot\varphi + \omega_\mathrm{nat}^2 \sin\varphi = 0 \; ,
    \label{eq:Thread_ODE_norm}
\end{equation}
where $Q$ is the quality factor \cite{reiss_parameter_2026}. Typical $Q$ factors in Thread-MMRs are around 1000, while singular values lie within the range of up to 3000 \cite{gleich_miniature_2023}.

\subsubsection{MMRs with jewel bearing}

Figure~\ref{fig:mmr-variants}\textit{b} depicts the proposed jewel-bearing-based MMR in a sectional view. The spherical rotor magnet rests on the concave surface of an industrial jewel with cup radius $R>r$. In jewel bearing terminology, the moving contact partner is called the tip. Here, this role is taken by the rotor magnet itself, which is why its spherical shape is mandatory for this type of MMR. Since the cup radius exceeds the tip radius, the two surfaces touch in a single point in the undeformed state. They transmit the load over the small area that results from their elastic deformation, treated in Section~\ref{ssec:characteristics-of-jewel-bearings}. A hard surface coating on the magnet pairs two hard materials at this contact, as is standard practice for the precision bearings referred to above \cite{krause_lager_2018}. Like in the Thread-MMR, the equilibrium position of the rotor on the jewel cup automatically corresponds to its desired position and orientation regarding the MMR's main axis and is held there by the magnetic forces.

In a first approximation, following the combined damping model of \citeauthor{kanj_design_2022}~\cite{kanj_design_2022}, the angle-dependent normal force at the jewel contact is here treated as constant, yielding a friction torque $T_\mathrm{f}$ that is independent of $\varphi$. Under this simplification, the equation of motion for the Jewel-MMR extends eq.~\eqref{eq:ODE} by both a viscous and a dry friction term to
\begin{equation}
    I \ddot\varphi + c \dot\varphi + m_\mathrm{r} B_0 \sin\varphi +
    \underbrace{\mathrm{sgn}\, \dot\varphi \ T_\mathrm{f}}_{T_\mathrm{Dry}} = 0 \;,
    \label{eq:Jewel_ODE_simp}
\end{equation}
with $\mathrm{sgn}$ being the sign function. Normalizing analogously to eq.~\eqref{eq:Thread_ODE_norm} yields
\begin{equation}
    \ddot\varphi + \frac{\omega_\mathrm{nat}}{Q} \dot\varphi + \omega_\mathrm{nat}^2
    \sin\varphi + \mathrm{sgn}\, \dot\varphi \ \frac{T_\mathrm{f}}{I} = 0 \; .
    \label{eq:Jewel_ODE_norm_simp}
\end{equation}
The function $\mathrm{sgn}\,\dot\varphi$ ensures that the constant dry friction torque always opposes the direction of motion, in contrast to the viscous term which scales linearly with $\dot\varphi$. Under the assumption that the dry friction term is dominant, the amplitude of a Jewel-MMR would decay approximately linearly rather than exponentially which has direct consequences for the choice of quality metrics.

\subsection{Quantifying MMR quality}
\label{ssec:quantifying_mmrs_quality}

\subsubsection*{Limitations of Standard Oscillator Metrics}

For damped harmonic oscillators, where viscous (linear) damping is the dominant loss mechanism, the quality factor $Q$ is a constant parameter and thus serves as the fundamental characteristic (e.g. \citeauthor{li_pivot_2025}~\cite{li_pivot_2025}). It is defined as the ratio of stored energy to energy dissipated per cycle and is inversely proportional to the viscous damping coefficient
\begin{equation}
    Q =\frac{I \, \omega_\mathrm{nat}}{c} \; .
\end{equation}
Since $Q$ is normalized to $\omega_\mathrm{nat}$ and $I$, it enables size-independent comparisons between different oscillator systems \cite{li_pivot_2025} and has already been used to characterize Thread-MMRs \cite{gleich_miniature_2023}. However, since dry (constant) friction is expected to be the dominant loss mechanism in Jewel-MMRs, and $Q$ is also defined above to be independent of $T_\mathrm{f}$, the quality factor loses its physical meaning as a universal figure of merit.

To nonetheless report a comparable parameter, an equivalent quality factor $\tilde Q$ is introduced for the Jewel-MMR. It is defined such that the amount of signal generated by a Jewel-MMR is compared to a reference oscillator with pure viscous damping. Let $A_\mathrm{Jewel}$ be the area under the envelope of the measured change in magnetic moment in the local $v$-channel $\hat{\dot m}_v(t)$ over the interval $t\in[0, T]$, expressed as
\begin{equation}
    A_\mathrm{Jewel} := \int_0^T \hat{\dot m}_v(t) \,\mathrm{d}t \;.
\end{equation}
Then, $A_\mathrm{Jewel}$ is defined to be equal to the area under an equivalent exponential function over the same interval such that
\begin{equation}
    A_\mathrm{Jewel} = \int_0^T \max\left(\hat{\dot m}_v\right) \cdot  \exp{\left( -\frac{\omega_\mathrm{nat}}{2\,\tilde{Q}_\mathrm{Jewel}}\, t \right) }\, \mathrm{d}t \;,
\end{equation}
which yields
\begin{equation}
    \tilde{Q}_\mathrm{Jewel} = \frac{A_\mathrm{Jewel} \, \omega_\mathrm{nat}}
    {2 \, \max\left(\hat{\dot m}_v\right)}
    \cdot
    \underbrace{\frac{1}{1 - \exp{\left(-\frac{\omega_\mathrm{nat}}{2\,\tilde{Q}_\mathrm{Jewel}}\, T\right)} }}_{=1 \mathrm{~for~} T\to\infty} \;,
    \label{eq:tilde-Q}
\end{equation}
that can be solved numerically, but also has an analytical solution for $T\to\infty$. This equivalent $Q$-factor is intended solely as a reference value, since it is dependent on the initial deflection angle $\varphi_\mathrm{max}$ through $\max(\hat{\dot m}_v)$ and the considered time interval $T$.

\subsubsection*{A Practical Evaluation Framework}

Given these limitations, this work adopts a two-tiered evaluation approach. The first tier concerns intra-variant comparisons, i.e. comparisons between different design parameter choices within the Jewel-MMR family. For these comparisons, the damping-related parameters provide a useful, direct, and sufficient ordering criterion. This holds especially for the friction torque $T_\mathrm{f}$. These parameters indicate the relative influence of design choices on energy dissipation. They cannot, however, be interpreted in absolute terms across variants.

For inter-variant comparisons between Thread-MMRs and Jewel-MMRs, the most meaningful and fair metric is the estimation precision achieved for the quantities from which sensing and tracking information is derived, under equal experimental conditions. Fair comparison requires both an equal magnet size and an equal natural frequency $\omega_\mathrm{nat}$. Both quantities govern the signal-to-noise ratio (SNR) of the measured response through the same mechanism: they increase the rate of change of the rotor magnet's magnetic moment, which constitutes the measured signal. The magnet size sets the magnitude of the magnetic moment, which scales with the cube of the radius $r$. The natural frequency sets how fast the moment varies in time. A larger moment or a higher frequency therefore both raise the signal amplitude and, with it, the SNR. Comparisons between MMRs that differ in either quantity are thus inherently unfair. Concretely, this refers to the statistical spread of a derived parameter over repeated measurements: for the natural frequency, the standard deviation of repeated frequency estimates; for the orientation, the mean angular deviation from a common reference pose. Since no independent optical or mechanical reference is available in the setup used here, this reference is taken from the measurements themselves, as detailed in Section \ref{ssec:analysis_of_measurement_data}, so that the metric characterizes repeatability rather than absolute accuracy. These metrics do not constitute a sensing or tracking task themselves, but they bound the precision any such application can attain with the respective resonator, and are therefore used for the final comparison in Section~\ref{sec:experiments}.

\subsubsection*{Manufacturing and Miniaturizability}

Beyond operational performance, the manufacturability of an MMR design constitutes a critical quality dimension, particularly given the small scales involved. Relevant aspects include the number of individual components, the complexity of assembly steps, the sensitivity of the design to manufacturing tolerances, and the extent to which the design is inherently self-aligning or otherwise robust by construction. A design that requires tight tolerances to achieve its nominal performance will inevitably exhibit high variability between specimens and may become impractical at smaller scales. These qualitative considerations therefore enter the final assessment of the MMR variants alongside the quantitative performance metrics introduced above, and will be revisited in Sections~\ref{ssec:types_of_manufacturing_errors} and~\ref{sec:experiments}.

As a first step toward this assessment, the following section derives the friction torque introduced by the jewel bearing as a function of its geometric parameters, establishing the theoretical foundation for the subsequent dimensioning of the Jewel-MMR.

\subsection{Characteristics of jewel bearings in MMRs}
\label{ssec:characteristics-of-jewel-bearings}

The mating surfaces of a jewel bearing's cup and tip ideally have a single contact point that results in a frictionless bearing. In practice, the hardness of the involved materials is limited; therefore, the force pushing the tip part into the cup material will result in a Hertzian compression of the contact partners \cite{krause_lager_2018}. The resulting deformation enlarges the actual contact area, creating a so-called pressure circle with radius $a$, which introduces the aforementioned friction torque $T_\mathrm{f}$ acting against the MMR's oscillation as of eq. \eqref{eq:Jewel_ODE_simp}. In the literature, $T_\mathrm{f}$ is estimated by integrating the infinitesimal friction torque contributions over the entire pressure circle, where $\mu$ is the coefficient of sliding friction and $F$ is the force acting perpendicular to the contact point \cite{krause_lager_2018}. It follows that
\begin{equation}
    T_\mathrm{f} = \frac{3}{16} \pi a \mu F
    \label{eq:tf}
\end{equation}
with the pressure circle of radius
\begin{equation}
    a  = \sqrt[3]{  \frac{3}{4}
                    \frac{1-\nu^2}{\frac{E_1 E_2}{E_1+E_2}}
                    \frac{rR}{R-r}
                    F} \; ,
    \label{eq:a}
\end{equation}
where the second fraction under the root contains all the material constants, and the third describes the dependence on the geometric parameters $r$ and $R$ of the jewel bearing as defined in Fig.~\ref{fig:mmr-variants}. $E_1$ and $E_2$ are the Young's moduli for the jewel and the magnet's coating material, respectively, and $\nu$ is the Poisson's ratio. It is estimated to be the same for both contact partners, for homogeneous materials, $\nu \approx 0.3$ \cite{krause_lager_2018}. It can be shown that the overall size of this bearing has no influence on the quality of the MMR's oscillation, as long as all geometric parameters are scaled equally. Jewel-MMRs can therefore be miniaturized without any size restrictions on the bearing. The proof of this claim and a detailed derivation of equations~\eqref{eq:tf} and~\eqref{eq:a} can be found in Appendices~\ref{sup:proof_size_independence} and~\ref{sup:derivation-friction-torque-jewel-bearing}.

So far, the friction torque $T_\mathrm{f}$ has been treated as constant. In reality, however, the normal force $F$ acting at the jewel contact varies with the deflection angle $\varphi$, since the magnetic attraction between stator and rotor depends on their relative orientation. For two dipoles separated by distance $d$, the magnitude of the magnetic force, which aligns with the main axis, evaluates to
\begin{equation}
    F_\mathrm{m} = \left|\boldsymbol{F}_\mathrm{m}\right| =
    \frac{4\pi B_\mathrm{r}^2}{3 \mu_0} \frac{r^6}{d^4} \,\cos\varphi
    =: \hat{F}_\mathrm{m} \cos\varphi \; ,
    \label{eq:fm(phi)}
\end{equation}
where $\hat{F}_\mathrm{m}$ collects all constant prefactors. Substituting eq.~\eqref{eq:fm(phi)} into eq.~\eqref{eq:tf} and absorbing all constant geometric and material parameters into a single amplitude $\hat{T}_\mathrm{f}$, the angle-dependent friction torque becomes
\begin{equation}
    T_\mathrm{f} = \frac{3\pi\mu}{16}
    \sqrt[3]{\frac{3}{4} \frac{1-\nu^2}{\frac{E_1 E_2}{E_1+E_2}} \frac{rR}{R-r}}
    \left[\hat{F}_\mathrm{m} \cos\varphi\right]^{\frac{4}{3}}
    =: \hat{T}_\mathrm{f} \cos^{\frac{4}{3}}\varphi \; .
    \label{eq:tf(phi)}
\end{equation}
The $\cos^{4/3}$ dependence shows that the friction torque is largest at the equilibrium position $\varphi = 0$ and decreases toward the turning points of the oscillation, where the magnetic attraction is weakest. Replacing the constant $T_\mathrm{f}$ in eq.~\eqref{eq:Jewel_ODE_simp} with the definition in eq.~\eqref{eq:tf(phi)} yields the final equations of motion for a Jewel-MMR in this work:
\begin{equation}
    I \ddot\varphi + c \dot\varphi + m_\mathrm{r} B_0 \sin\varphi +
    \mathrm{sgn}\, \dot\varphi \ \hat{T}_\mathrm{f} \cos^{\frac{4}{3}}\varphi = 0
    \label{eq:Jewel_ODE}
\end{equation}
and in normalized form
\begin{equation}
    \ddot\varphi + \frac{\omega_\mathrm{nat}}{Q} \dot\varphi +
    \omega_\mathrm{nat}^2 \sin\varphi +
    \mathrm{sgn}\, \dot\varphi \ \frac{\hat{T}_\mathrm{f}}{I}
    \cos^{\frac{4}{3}}\varphi = 0 \; .
    \label{eq:Jewel_ODE_norm}
\end{equation}
They collapse to the simplified eqs.~\eqref{eq:Jewel_ODE_simp} and~\eqref{eq:Jewel_ODE_norm_simp} in the limit $\varphi \to 0$. It directly follows that the effective friction torque
of a Jewel-MMR can be reduced just by increasing the deflection angle. This is also favorable in the general case of MMRs, since higher deflection angles yield higher angular velocities $\dot \varphi$ and therefore induce a signal of greater amplitude improving the SNR. This treatment of $T_\mathrm{f}$ rests on a set of assumptions that delimit its scope: the Hertzian contact is evaluated quasi-statically, i.e., the pressure circle is assumed to adapt instantaneously to the momentary normal force without inertial, viscoelastic, or impact effects, both mating surfaces are taken as ideally smooth, and neither wear nor foreign particles such as dust or debris are represented.

With the aim of finding suitable geometric parameters for the concave jewel cup, it can first be determined that $R$ cannot be negative for the present case, since that would result in a convex jewel surface\footnote{By definition, a negative spherical shell radius corresponds to a spherical surface with the same radius in terms of absolute value (see Appendix \ref{sup:derivation-friction-torque-jewel-bearing}).} that forms together with the tip an unstable contact point with no stable equilibrium position. Keeping all material-related parameters and the contact force constant, it can be derived from eqs. \eqref{eq:tf} and \eqref{eq:a} that for a given tip radius $r$ the jewel cup with the lowest friction torque would have a flat surface ($R\rightarrow\infty$). Introducing $R=\varkappa\cdot r$, which expresses the cup radius $R$ as a multiple of the tip radius $r$, and normalizing the resulting friction torque $T_\mathrm{f}(\varkappa)$ to the optimal value $T_\mathrm{f},\vert_{R\to\infty}$ yields
\begin{equation}
    \frac{\hat{T}_\mathrm{f}(\varkappa)}{\hat{T}_\mathrm{f}\,\vert_{R\to\infty}} =: \Lambda(\varkappa) = \sqrt[3]{\frac{\varkappa}{\varkappa-1}} \;.
\end{equation}

\begin{figure}[pos=t]
\centering
\includegraphics[width=\columnwidth]{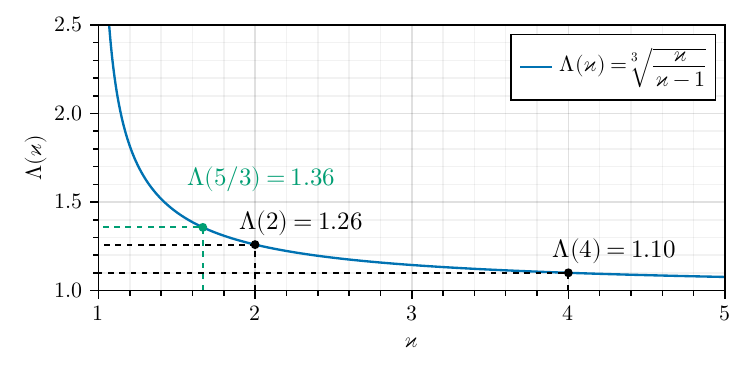}
\vspace{-1cm}
\caption{The auxiliary function $\Lambda(\varkappa)$ describes the proportional increase in friction torque of the Hertzian contact for cup-shaped jewels with radius $R$ and a sphere with radius $r$ relative to a flat jewel ($R\to\infty$) as a function of $\varkappa$ with $\varkappa=\frac{R}{r}$. The point marked in green represents the value for the Jewel-MMR that is used for the experiments.}
\label{fig:simulation_lambda(t)}
\end{figure}

Figure \ref{fig:simulation_lambda(t)} plots this newly defined auxiliary function $\Lambda(\varkappa)$. With, for example, $\Lambda(2) \approx 1.26$ and $\Lambda(4) \approx 1.10$, it can be seen that a cup radius of only twice or four times the tip radius is already sufficient to get close to the geometrical optimum, while having the advantages of a more stable equilibrium position due to the greater curvature. As the manufacturing errors of MMRs and their influence on the oscillation quality are explored next, it will be shown that these are relevant considerations during the design process.

\subsection{Types of manufacturing errors in MMRs}
\label{ssec:types_of_manufacturing_errors}

\begin{figure}[pos=ht]
\centering
\includegraphics[width=\columnwidth]{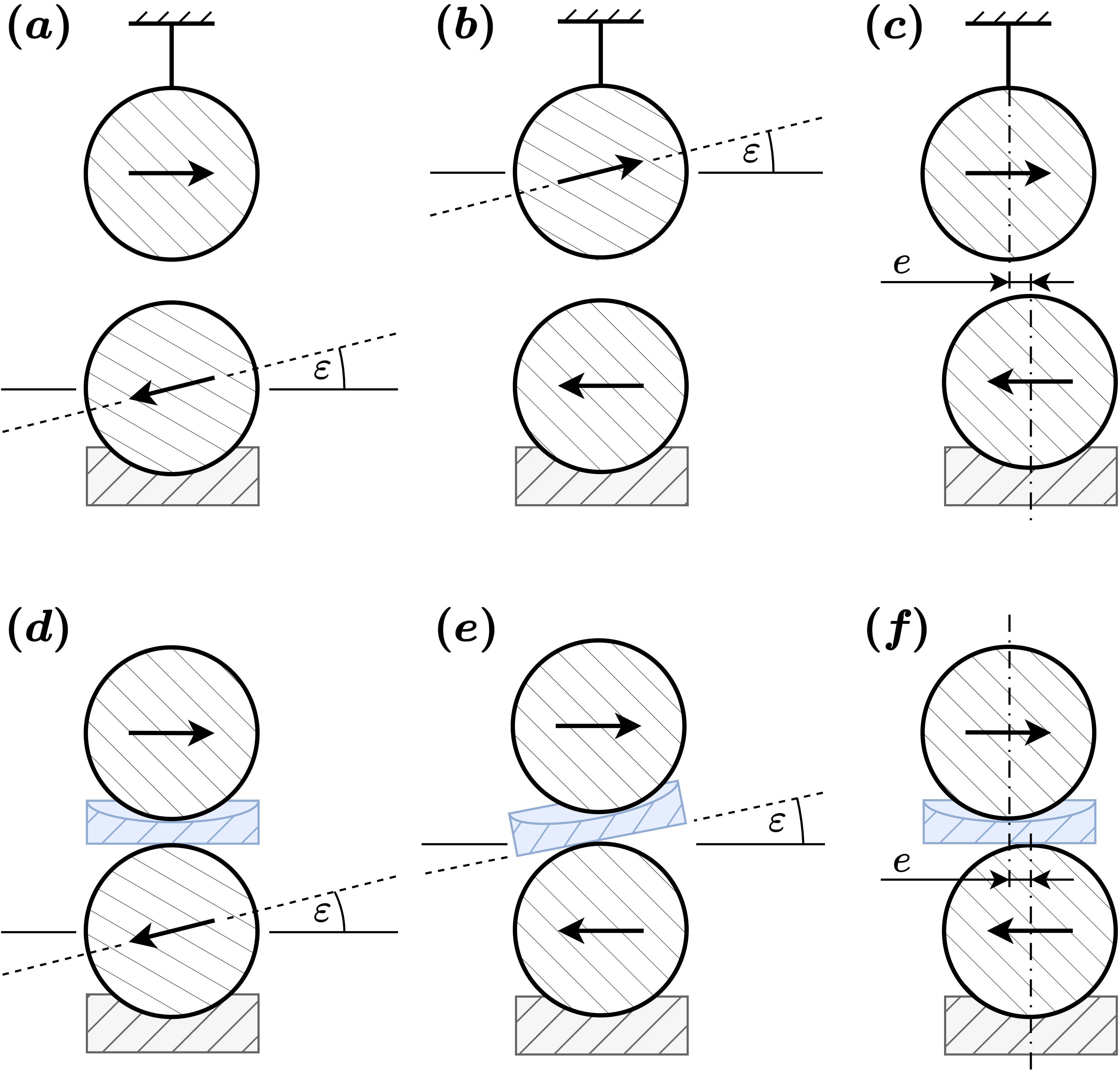}
\caption{Complete overview of positioning and alignment defects during manufacturing for Thread- and Jewel-MMRs in sectional view. The spherical stator and rotor magnets are considered ideal. Arrows indicate their initial orientation before the stator causes a change in the rotor. For a Thread-MMR, the error patterns are \textit{(a)} tilted stator, \textit{(b)} tilted rotor, and \textit{(c)} radial deviation. For the Jewel-MMR, they are \textit{(d)} tilted stator, \textit{(e)} tilted jewel, and \textit{(f)} radial deviation. Tilt errors are specified by the angle $\varepsilon$, radial deviations with $e$.}
\label{fig:diagram-manufacturing-defects-schematic}
\end{figure}

For the Thread- and Jewel-MMR, Fig.~\ref{fig:diagram-manufacturing-defects-schematic} depicts all possible manufacturing defects regarding positioning and alignment that would cause deviations from the ideal MMR. Each identified defect is shown isolated. An actual MMR would have all defects simultaneously in superposition. To further simplify the considerations for now, the visualization excludes the impact of each defect on the rotor's new equilibrium position, and all magnets are assumed to be ideal, that is, uniformly magnetized. Also, deviations that only change the distance between the magnets are not considered a defect, as they result only in a variation of natural frequency in an otherwise ideal MMR.

Three manufacturing errors can be identified for Thread-MMRs: a tilt of the stator (Fig.~\ref{fig:diagram-manufacturing-defects-schematic}\textit{a}) or rotor (Fig.~\ref{fig:diagram-manufacturing-defects-schematic}\textit{b}) by an error angle $\varepsilon$, or a radial deviation $e$ of misaligned axes (Fig.~\ref{fig:diagram-manufacturing-defects-schematic}\textit{c}). Even with cylindrical magnets, whose flat surfaces serve as geometric markers for correct vertical orientation, a tilt error can occur if the thread is not attached centrally to the magnet. Since for Jewel-MMRs the rotor magnet self-aligns and is not attached to anything, only a misalignment of the stator as in Fig.~\ref{fig:diagram-manufacturing-defects-schematic}\textit{d} constitutes a manufacturing error. Instead, Fig.~\ref{fig:diagram-manufacturing-defects-schematic}\textit{e} shows a tilt of the jewel with respect to its horizontal position, and Fig.~\ref{fig:diagram-manufacturing-defects-schematic}\textit{f} depicts a radial deviation between the jewel's and the stator's axis. For the special case of a flat jewel, defect \textit{(f)} would be not applicable and errors \textit{(d)} and \textit{(e)} would describe the same behavior, yielding only one source of error. But even for the general case of a jewel cup, defect pattern \textit{(d)} could be viewed as a superposition of \textit{(e)} and \textit{(f)}, still reducing the number of error sources by one.

All error patterns in Fig.~\ref{fig:diagram-manufacturing-defects-schematic} are shown initially before the magnetic forces are applied and the rotors would find their equilibrium position. It can be shown that, e.g., tilt errors \textit{(a)} and \textit{(b)} of the Thread-MMR in Fig.~\ref{fig:diagram-manufacturing-defects-schematic} yield new, visually similar equilibrium positions for the rotor (see Appendix \ref{sup:additional_manufacturing_defects}). This, in turn, may also lead to equivalent oscillatory behavior. The decisive difference compared to the Jewel-MMR, however, is that each source of tilt error occurs independently twice during the Thread-MMR's assembly. Namely, once for the placement of the stator and once for the thread-rotor attachment. The self-aligning property of the Jewel-MMR reduces this to a single source of possible misalignment.

Despite the apparent diversity of defect types, all identified errors share a common underlying mechanism of introducing asymmetries in the design. These asymmetries force the rotor to oscillate about an axis that is neither perpendicular to its magnetization direction nor does it pass through both magnets' centers. This includes, incidentally, the effect of gravity, which introduces an equivalent static tilt for any non-vertical main axis orientation. The tilt error of the stator is studied in the following section as the primary defect. It occurs in both variants, and it represents a major uncertainty during assembly: since the magnets are spherical and their magnetization direction carries no visible geometric marker, a tilt error is very likely, whereas lateral displacements can be controlled far more precisely through the geometry of the housing. In Appendix \ref{sup:additional_manufacturing_defects}, the remaining defects are shown to yield simulation results similar to those of the stator tilt. The only exception is the lateral displacement, whose equivalent perpendicular force vanishes in the limiting cases of a flat jewel ($R \to \infty$) or infinitely long thread ($l \to \infty$).

\subsection{Relationship between manufacturing errors and side mode susceptibility}
\label{ssec:manufacturing_errors_side_modes}

\begin{figure}[pos=ht]
    \centering
    \includegraphics[width=\columnwidth]{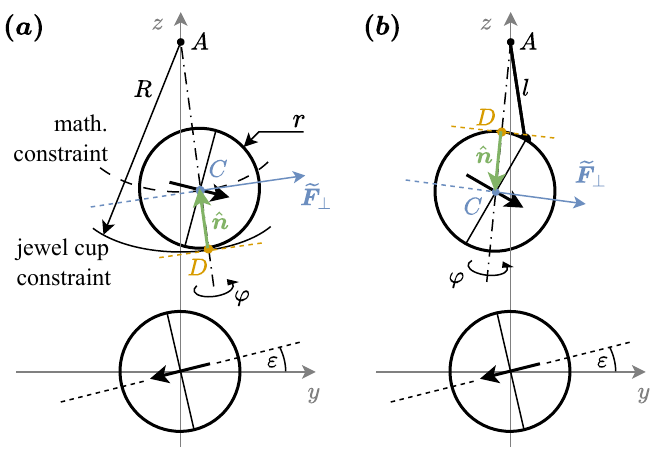}
    \caption{Schematic of a \textit{(a)} Jewel- and a \textit{(b)} Thread-MMR with a stator manufacturing defect. Stator and rotor magnets are shown as spheres projected onto the $yz$-plane. Per MMR the stator is tilted by an angle $\varepsilon$ about the $x$-axis that leads to a change in the initial equilibrium position $C$ of the respective rotor. Arrows mark the orientation of magnetization. The jewel bearing is modeled with a spherical plane of radius $R$, the thread bearing with a rigid connection of length $l$. Both have their anchor point in $A$. For subsequent considerations, the MMR is deflected by angle $\varphi$ about its rotational axis for the new equilibrium position which is defined through points $A$ and $C$. For illustrative purposes, examples of the 3-dimensional output measure $\widetilde{F}_\perp$, representing the susceptibility to side mode excitation, are also displayed as projections onto the $yz$-plane.}
    \label{fig:math_description_angle_defects}
\end{figure}

As established above, the rotor of an ideal MMR under ideal excitation only experiences a magnetic torque about and a magnetic force along its axis of rotation, leaving the side modes unexcited. Any force $\boldsymbol{F}_\perp$ or moment $\boldsymbol{T}_\perp$ acting on the rotor perpendicular to this axis therefore represents the necessary condition for side mode excitation. For the following simplified static investigations, the magnitude of an equivalent perpendicular force $\widetilde{F}_\perp$ serves as a simple measure of side mode susceptibility. To collapse the perpendicular force and moment into a single quantity, the force–torque pair acting on the rotor is reduced to the virtual bearing point $D$, i.e., the point at which the axis of rotation pierces the rotor's surface, located a distance $r$ (the magnet radius, and thus the effective lever arm, see Fig. \ref{fig:math_description_angle_defects}) from its center. With $\hat{\boldsymbol{n}}$ the unit vector along the axis of rotation pointing from $D$ to the magnet center, the moment about $D$ reads $\boldsymbol{M}_D = \boldsymbol{T} + r\,\hat{\boldsymbol{n}} \times \boldsymbol{F}$, and only its perpendicular component $\boldsymbol{M}_{D,\perp} = \boldsymbol{M}_D - (\boldsymbol{M}_D \cdot \hat{\boldsymbol{n}})\,\hat{\boldsymbol{n}}$ can tilt the rotor. Thereby this moment feeds the side modes, whereas the component along $\hat{\boldsymbol{n}}$ merely adds to the drive moment of the intended working mode. The equivalent perpendicular force is thus defined as
\begin{equation}
    \widetilde{F}_\perp := \left\vert \widetilde{\boldsymbol{F}}_\perp \right\vert = \frac{\left\vert \boldsymbol{M}_{D,\perp} \right\vert}{r} \;.%
\end{equation}
Referencing the perpendicular moment to $D$ rather than to the actual thread contact point guarantees that $\widetilde{F}_\perp$ is taken strictly perpendicular to the axis of rotation, so that no restoring moment of the working mode leaks into the measure; for the jewel bearing $D$ coincides with the physical contact point. This approach intentionally neglects the complex motion patterns of such an asymmetric system and serves here merely as a rough estimate of the underlying trends. A dynamic analysis of the actual motion patterns lies beyond the scope of this work.

Figure~\ref{fig:math_description_angle_defects} depicts the Jewel- and the Thread-MMR with a tilted stator defect schematically. Each MMR is modeled as a system of two interacting magnetic dipoles with simplified bearing constraints while neglecting any influence by friction or gravity. For illustrative purposes, examples of $\widetilde{F}_\perp$ are also displayed as projections onto the $yz$-plane. The jewel bearing cup is modeled as a fixed spherical plane with radius $R$ on which the rotor dipole is free to move. Mathematically, this is achieved through a magnetic dipole which is bound to a sphere with radius $R-r$, that is, the cup radius reduced by the sphere radius $r$. For the Thread-MMR, the bearing is modeled as a rigid connection between its anchor point $A$ on the $z$-axis and the rotor's neutral line. It results in the degrees of freedom similar to those of a double pendulum.

For the static problem analysis, the output variable $\widetilde{F}_\perp$ is computed per MMR model according to the following procedure:
\begin{enumerate}
    \item The stator dipole is tilted by the error angle $\varepsilon$ about the $x$-axis, which is perpendicular to the initial main axis.
    \item A minimization algorithm finds the rotor dipole's new equilibrium position taking into account the bearing constraints.
    \item The rotational axis of the main mode is redefined through the anchor point $A$ and the rotor's center point~$C$.
    \item The rotor is deflected from its equilibrium position by an angle $\varphi$ about the new axis of rotation.
    \item The force and torque components, acting perpendicular to the main rotational axis, are extracted by solving the statics problem.
    \item The equivalent perpendicular force magnitude is calculated.
\end{enumerate}

\begin{figure*}[pos=ht]
    \centering
    \includegraphics[width=\textwidth]{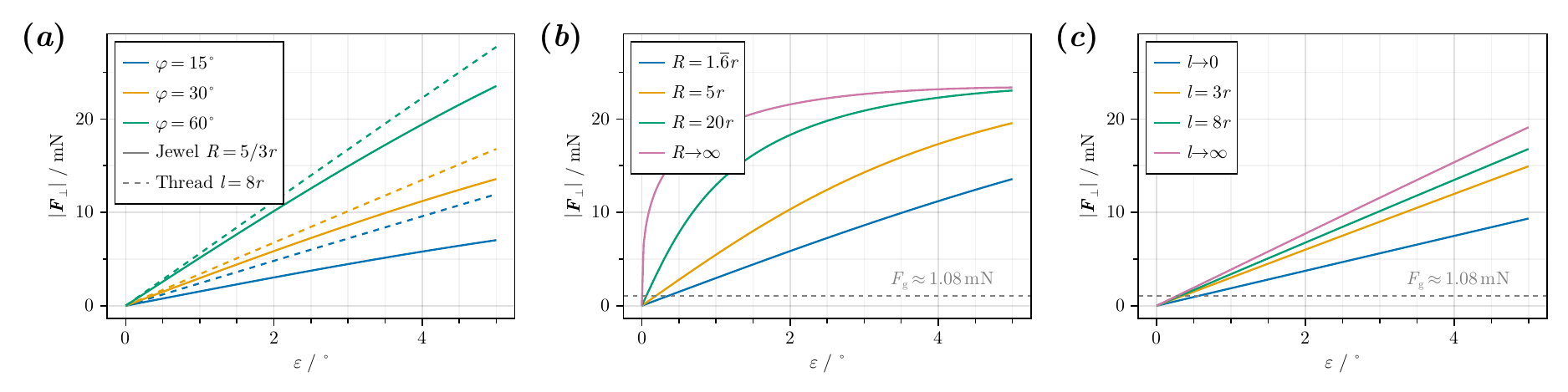}
    \plotspace
    \caption{Simulation results of the equivalent perpendicular force magnitude $\widetilde{F}_\perp$ in MMRs with tilted stator defect of error angle~$\varepsilon$. Subplot \textit{(a)} shows multiple deflection angles $\varphi$ at the cup radius $R=\frac{5}{3}r$ for the Jewel-MMR and at the thread length $l=8r$ for the Thread-MMR. Subplots \textit{(b)} and \textit{(c)} show results with a fixed deflection angle $\varphi=30^\circ$ for multiple cup radii and thread lengths respectively. The magnets of both MMRs have a radius of $r=1.5$~mm, density of $\rho=7800$~kg~m$^{-3}$, remanent magnetization of $B_\mathrm{r}=1.3$~T and an initial distance of $d=4.7$~mm.}
    \label{fig:F_tan_over_epsilon}
\end{figure*}

The plots in Fig.~\ref{fig:F_tan_over_epsilon} show the resulting equivalent perpendicular force as a function of the error angle $\varepsilon$ for the tilted stator magnet. In Fig.~\ref{fig:F_tan_over_epsilon}\textit{a} several deflection angles $\varphi$ for both MMRs are displayed next to each other. To match subsequent measurements, a cup radius of $R=\frac{5}{3}r$ and a thread length of $l=8r$ were chosen. Figures~\ref{fig:F_tan_over_epsilon}\textit{b} and \ref{fig:F_tan_over_epsilon}\textit{c} explore the dependencies on the design parameters $R$ and $l$ at a fixed deflection angle of $\varphi=30^\circ$ respectively.

Four observations are central. First, for a perfectly manufactured MMR, i.e. $\varepsilon = 0$, $\widetilde{F}_\perp$ vanishes identically, following the considerations in eqs.~\eqref{eq:T_m} and~\eqref{eq:T_m_2} that side modes are not excited under ideal conditions. However, even small manufacturing errors lead to non-negligible equivalent perpendicular forces in the order of magnitude of gravity ($F_\mathrm{g}\approx1.08$~mN). In preliminary experiments, it was observed that gravity alone visibly increased side mode oscillations in the measurement signal when the MMR was mounted with its main axis horizontally aligned. Taking $F_\mathrm{g}$ as an empirically anchored reference for a perturbation that measurably excites side modes, the curves in Fig.~\ref{fig:F_tan_over_epsilon}\textit{a} reach this magnitude already at error angles up to $\varepsilon\approx0.7^\circ$, depending on the deflection angle and the bearing variant. Manufacturing tolerances therefore would have to be kept well below one degree for the perpendicular force to stay below a level that is known to excite side modes in practice. Second, larger deflection angles $\varphi$ are associated with higher $\widetilde{F}_\perp$ as in Fig.~\ref{fig:F_tan_over_epsilon}\textit{a}, suggesting that in this case the risk of side mode excitation increases. This behavior is also consistent with the observations from the subsequent experiments. Consistent with this, a more precisely manufactured MMR would be expected to tolerate higher oscillation angles and thereby a higher achievable SNR. Third, larger cup radii also increase $\widetilde{F}_\perp$ according to Fig.~\ref{fig:F_tan_over_epsilon}\textit{b}. For very large cup radii $R\gg r$ the relationship even becomes strongly nonlinear: the force grows rapidly even for small $\varepsilon$ and then flattens, indicating that the first few degrees of misalignment are the most critical. This is also consistent with preliminary experiments using flat jewels, with which practically no stable central equilibrium position could be achieved for the rotor. Fourth, a longer thread also seems to increase the equivalent perpendicular force in Thread-MMRs slightly. Regardless, since the adhesive, which is typically used for their assembly, partially creeps into the thread and stiffens it, a longer thread was chosen in this work to ensure low damping of the oscillation.

In contrast to the friction torque analysis in Section~\ref{ssec:characteristics-of-jewel-bearings}, where a larger $R$ was found to reduce $T_\mathrm{f}$ for the Jewel-MMR, these findings suggest that a smaller cup radius is beneficial. A tighter cup therefore appears to provide a stronger geometric centering effect that partially compensates for any misalignment of the stator, which reduces $\widetilde{F}_\perp$ in the model. For the limiting case $R \rightarrow \infty$ (flat jewel), the centering effect vanishes entirely due to the rotational symmetry of the flat surface, and the equivalent perpendicular force reaches its maximum for a given set of $\varepsilon$ and $\varphi$. Minimizing friction torque on one hand and maximizing side mode resilience on the other are the two competing requirements that constitute the central design trade-off for the Jewel-MMR cup geometry which is taken into account in the following section.

\section{Experiments}
\label{sec:experiments}

The conducted experiments serve several purposes. The primary goal is to have a fair, inter-variant comparison between the performance of a Jewel-MMR and that of a standard Thread-MMR, taking into account the two usage scenarios of tracking and sensing. Specifically, for tracking, the aim is to achieve the smallest possible error in the orientation estimate, and for sensing, the most precise estimate of the natural frequency. Both resonators remain at the same fixed pose at the center of the coil arrangement throughout. Since no measurand is applied and no position is estimated, the experiments address the quantities on which sensing and tracking rest, not the applications themselves. This restriction is deliberate and does not narrow the comparison. As eqs.~\eqref{eq:u(t,p,R)} and~\eqref{eq:projection} show, the pose dependence of the received signal resides entirely in the coil sensitivity $\boldsymbol{P}(\boldsymbol{p}_\mathrm{rotor})$ and the orientation $\boldsymbol{R}$, which are properties of the detection platform and of the geometry, whereas the bearing enters the signal exclusively through $\varphi(t)$. At equal magnet size and equal natural frequency, a measurement at a single fixed pose therefore isolates precisely the degree of freedom in which the two variants differ. What changes across a working volume is, to first order, the signal-to-noise ratio, and the comparison is parameterized over that quantity in Section~\ref{ssec:primary_results}. Additionally, the newly developed Jewel-MMR model in \eqref{eq:Jewel_ODE} is to be validated as a secondary goal.

\subsection{Materials and sizing of the MMRs}
\begin{figure}[pos=t]
    \centering
    \includegraphics[width=\columnwidth]{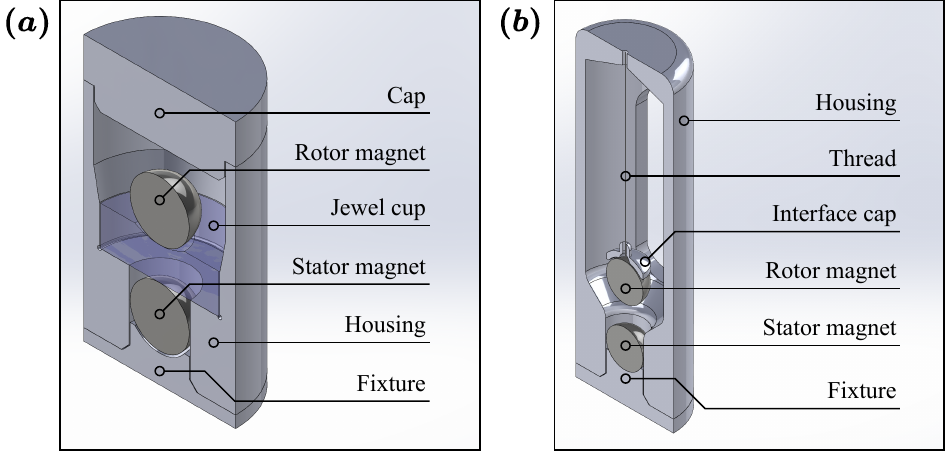}
    \caption{Labeled CAD models' section view for \textit{(a)} the Jewel-MMR and \textit{(b)} the Thread-MMR.}
    \label{fig:mmr-cad-models}
\end{figure}

\begin{figure}[pos=t]
    \centering
    \includegraphics[width=.7\columnwidth]{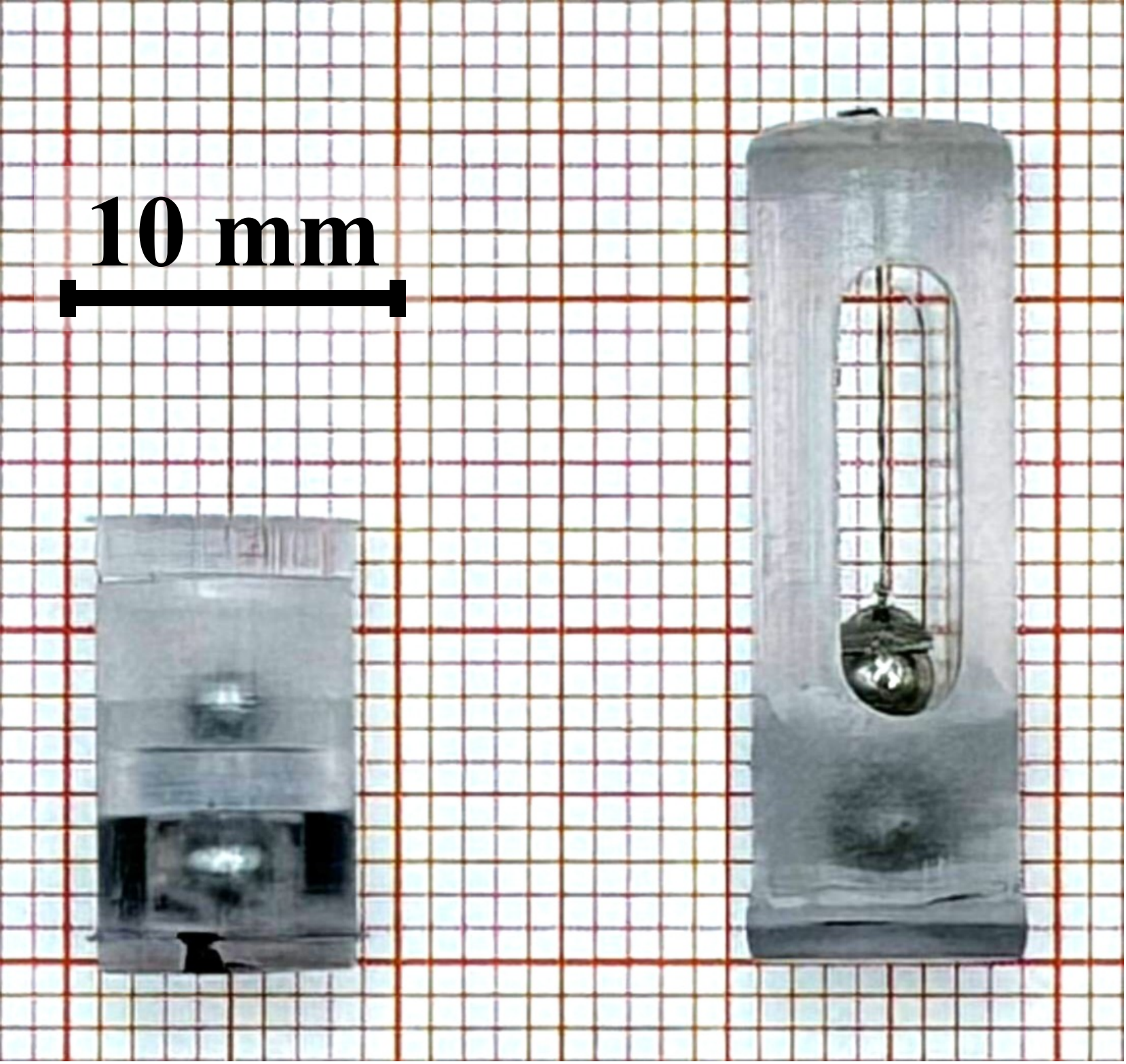}
    \caption{Picture of the assembled Jewel-MMR on the left and the Thread-MMR on the right side.}
    \label{fig:mmr-picture}
\end{figure}

Figures \ref{fig:mmr-cad-models} and \ref{fig:mmr-picture} show the section view of the CAD models and finished assemblies of both Jewel- and Thread-MMR, respectively. Their stators and rotors are all spherical sintered Neodymium (NdFeB) magnets, N42 grade, with a Chrome-plated (NiCuNiCr) surface\footnote{K-3-NI-N42 magnets from MTS Magnete $\vert$ Obere Wiesen 4, 78166 Donaueschingen, Germany $\vert$ \href{https://www.mtsmagnete.de/kugeln/kugeln-1mm-9mm/kugelmagnet-aus-neodym-3-mm-verchromt-n42/a-3124}{www.mtsmagnete.de}}. They have an outer diameter of $3 \mathrm{~mm}$ ($r=1.5\mathrm{~mm}$) and a remanent magnetization of $B_\mathrm{r}\approx1.3$ T. This yields a magnetic moment of
\begin{equation}
    m_\mathrm{r} = V \cdot \frac{B_\mathrm{r}}{\mu_0} = \frac{4\pi r^3 B_\mathrm{r}}{3 \mu_0} \approx 14.6 \mathrm{~mAm}^2 \; .
    \label{eq:m_r}
\end{equation}

Taking into account the previous findings regarding the cup radius $R$ and available jewel sizes on the market, a cylindrical sapphire glass jewel\footnote{SL-K6x2.5R Saphir Kalottenstein from SITUS Technicals GmbH $\vert$ Vohwinkeler Str. 58, 42329 Wuppertal, Germany} with $R=\frac{5}{3}r=2.5\mathrm{~mm}$ was selected for the Jewel-MMR. A tight stack of the magnets with the jewel as a separator, as shown in Fig.~\ref{fig:mmr-cad-models}\textit{a}, yields a center-to-center distance of $d\approx4.7\mathrm{~mm}$ between the stator and the rotor magnet. With eq. \eqref{eq:omega_nat}, follows for the MMR's natural frequency
\begin{equation}
    f_\mathrm{nat} = \frac{\omega_\mathrm{nat}}{2\pi} = \frac{1}{2\pi} \sqrt{\frac{5B_\mathrm{r}^2r}{6 \rho d^3 \mu_0}} \approx 229.3 \;\mathrm{Hz} \; .
    \label{eq:fnat}
\end{equation}
Using standard material parameters\footnote{$E_\mathrm{Sapphire}=460\mathrm{~GPa}\,,\; E_\mathrm{Chromium}=279 \mathrm{~GPa}\,,\; \mu=0.12\,,\; \nu = 0.3$} for an estimate of $\hat{F}_\mathrm{m}$ and $\hat{T}_\mathrm{f}$, with equations \eqref{eq:fm(phi)} and \eqref{eq:tf(phi)} it follows that $\hat{F}_\mathrm{m} \approx 131.5 \mathrm{~mN}$ and $\hat{T}_{\mathrm{f}} \approx 115.9 \mathrm{~nNm}$,
and for the friction term in the normalized ODE problem as in eq. \eqref{eq:Jewel_ODE_norm}
\begin{equation}
    \frac{\hat{T}_\mathrm{f}}{I} \approx 1168 \mathrm{~rad s}^{-2} \; .
    \label{eq:TfoverI}
\end{equation}

For the thread, a commonly used material in MMRs is High-Modulus Polyethylene (HMPE) fibers \cite{gleich_miniature_2023,faltinath_natural_2025}, often recognized solely under the brand name \emph{Dyneema}. For the MMR thread in this work, one manually isolated strand of a four-strand braided HMPE fishing line\footnote{WFT Round Dynabraid G4 grün 11 kg, \O 0.10 mm from World Fishing Tackle GmbH \& Co. KG $\vert$ Kaldenhausener Str. 41, 47802 Krefeld, Germany} is used. To ensure a fair comparison between the Jewel- and the Thread-MMR, a similar change in magnetic moment is targeted. Assuming that both rotor magnets have the same magnetic moment, MMRs with the same natural frequency should produce the same signal amplitude, in this work also referred to as signal strength. Instead of aiming for the same center-to-center distance for the Thread-MMR's magnets, the thread is trimmed in length during a preliminary experiment until the Thread-MMR's oscillation frequency, derived from a FFT analysis, roughly matches the Jewel-MMR's natural one. This way, the assembly errors can be reduced to the tolerances in the magnets' magnetization. Also, the resulting MMRs should roughly have the same stability, since their magnetic attraction force $\hat{F}_\mathrm{m}$ which tries to keep the rotor in its desired equilibrium position, is similar. The design thread length derived from the CAD model evaluates to $l_\mathrm{Thread}\approx8r\approx12\mathrm{~mm}$. An interface cap fixes the correct positioning of the thread onto the rotor magnet during assembly. The resulting increase in inertia is neglected, since any disadvantages in SNR are also offset by the frequency matching.

The components for the housings, fixtures and the interface cap are 3D-printed from a translucent photopolymer\footnote{Form 4 printer and Clear Resin V5 from Formlabs Inc. $\vert$ 35 Medford Street, Somerville, Massachusetts 02145, USA $\vert$ \url{www.formlabs.com}} and assembled with cyanoacrylate. Before closing the Jewel MMR with its cap, its rotor magnet is thoroughly cleaned with acetone to minimize the influence of foreign particles on the bearing. 

\subsection{Test Environment}
\begin{figure}[pos=t]
    \centering
    \includegraphics[width=\columnwidth]{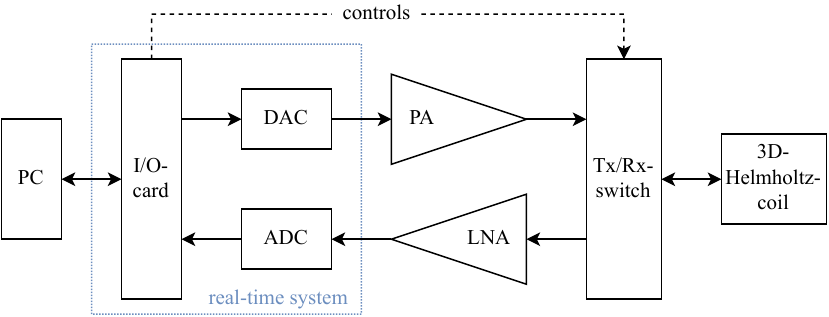}
    \caption{Schematic diagram of the detection platform consisting of an analog-to-digital-converter (ADC), a digital-to-analog-converter (DAC), an input/output-card (I/O-card), a low-noise amplifier (LNA), an audio class-D power amplifier~(PA), a transmit/receive-switch (Tx/Rx-switch), a PC and a 3D Helmholtz coil arrangement.}
    \label{fig:detection_platform}
\end{figure}

\begin{figure}[pos=t]
    \centering
    \includegraphics[width=.8\columnwidth]{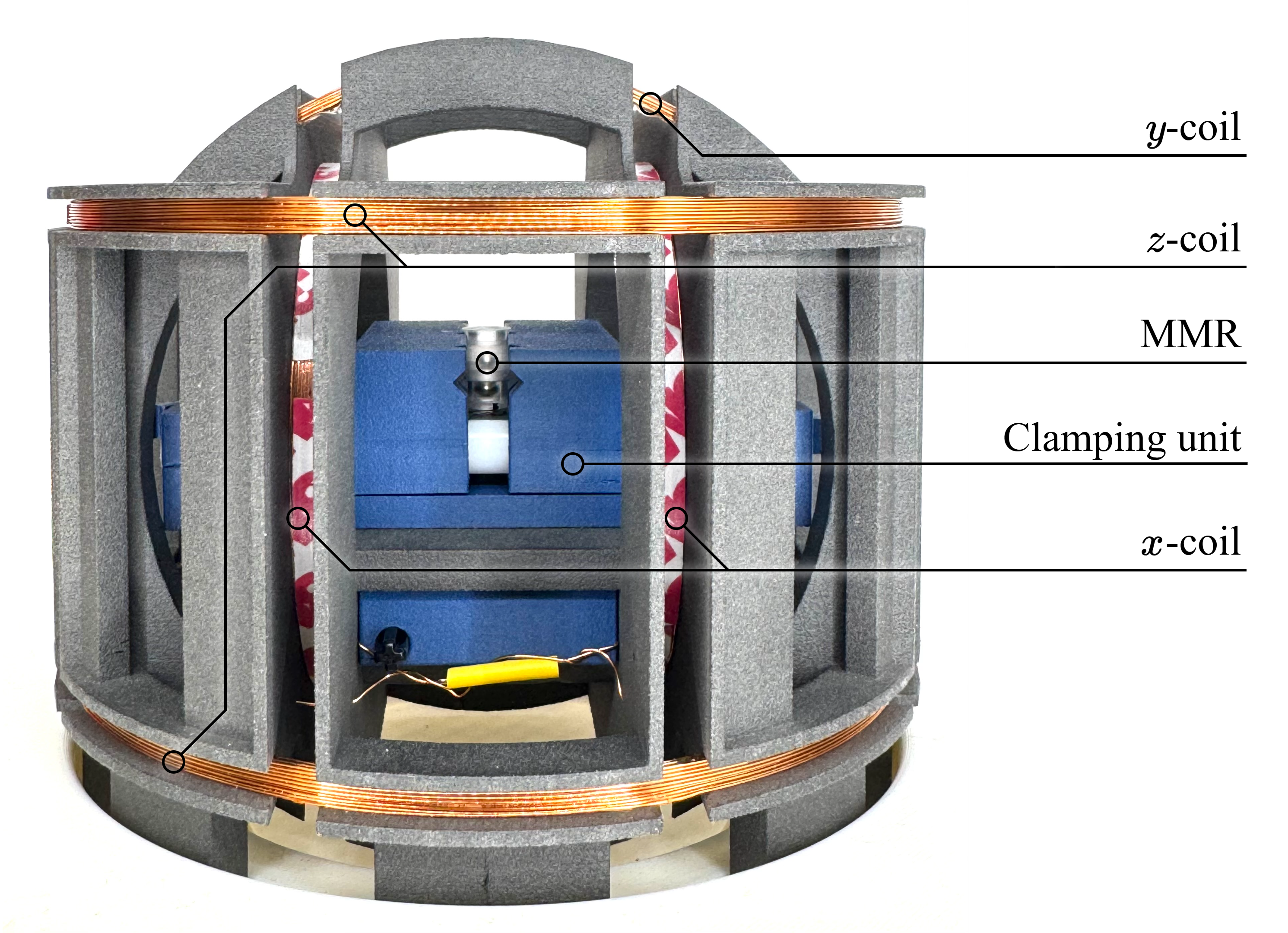}
    \caption{Photo of the 3D Helmholtz coil arrangement (dark gray) surrounding the clamping unit (blue), which, as an example, fixates a Jewel-MMR at the center of the coils.}
    \label{fig:coil-array-picture}
\end{figure}

Figure \ref{fig:detection_platform} depicts the schematic configuration of the experiment's detection platform. At the center of the detection system sits a clamping unit that can fixate an MMR in the measurement setup. In all experiments, the MMRs are oriented with their main axis collinear to the global $z$-axis and with their $v$-axis pointing in the $x$-direction. As depicted in Fig. \ref{fig:coil-array-picture}, it is surrounded by a 3D Helmholtz coil arrangement capable of generating spatially homogeneous, time-variant magnetic fields in any direction. By definition, the three coil dimensions correspond to the axes of the world coordinate system. A custom Julia framework \cite{reiss_parameter_2026} running on a consumer-grade computer\footnote{MINIS FORUM UM890 Pro Mini PC - AMD AI Ryzen 9 8945HS, 8~C/16~T, Radeon 780M GPU, 32GB DDR5 RAM, 1TB M.2 SSD} controls the experiments. From a given set of measurement parameters, further described in the following section, it calculates the excitation sequence. This is sent to a real-time system that generates the desired analog output. The signal is then amplified by a class-D audio amplifier and switched onto the coil arrangement. The resulting magnetic field excites the MMR. After the transmission, the same coils act as receivers for the MMR's response signal. A dedicated Tx/Rx-switch, also controlled by the I/O-card, disconnects the coils from the power amplifier and connects them to the receive chain. A low-noise amplifier prepares the signals for each channel before being digitized by the I/O-card with a sampling frequency of $\frac{125\mathrm{~MHz}}{2048}\approx 61$~kHz. Finally, the real-time system handles the communication with the asynchronously working program, buffering the measured data. \cite{mohn_transmit_2025}

\subsection{Measurement Sequence}
\label{ssec:measurement_sequence}

Table~\ref{tab:measurement_parameter} gives an overview of the parameters of the measurement sequence. They were derived through preliminary experiments. After sinusoidal excitation with the analytically found $f_\mathrm{nat}$ in \eqref{eq:fnat}, an initial frequency analysis over the Jewel-MMR's response provided an estimate for its actual natural frequency of approximately $212 \mathrm{~Hz}$. As described in the previous section, the Thread-MMR was then roughly trimmed to the same frequency. Overall, a short excitation time $t_\mathrm{TX}$ was chosen to reduce the phase error towards the sinusoidal excitation signal, which stems from the anharmonic nature of the MMRs. With $t_\mathrm{TX}=47\mathrm{~ms}$ the system excites the MMRs with around 10 oscillation periods. The amplitude for the sinusoidal magnetic excitation field $\hat{B}_\mathrm{TX}$ was then increased in steps of $10$~µT until the received signal showed signs of side mode oscillations, visible as signals in the detection system's $z$-channel. For the Jewel-MMR, this effect occurred at $\hat{B}_\mathrm{max} \approx 610$~µT, for the Thread-MMR at around $270$~µT. The final $\hat{B}_\mathrm{TX}$ parameter was then set to 80\% of each found maximum field strength $\hat{B}_\mathrm{max}$ for robust excitation within the MMRs' stable range of main mode excitation. The receive time $t_\mathrm{RX}$ takes into account the different ring-down behaviors of the MMRs, to ensure that the next measurement sequence acts upon a fully settled MMR and the same initial conditions are applicable.

\begin{table}[pos=h]
    \renewcommand{\arraystretch}{1.3}
    \centering
    \caption{Parameters of the measurement sequences}
    \label{tab:measurement_parameter}
    \begin{tabular}{l|cccc}
        MMR  & $f_\mathrm{TX} \; / \; \mathrm{Hz}$ & $t_\mathrm{TX} \; / \; \mathrm{s}$ & $\hat{B}_\mathrm{TX} \; / \; \text{µ}\mathrm{T}$ & $t_\mathrm{RX} \; / \; \mathrm{s}$ \\[1pt]
        \hline
        Jewel & 212 & 0.047 & 490 & 2.0 \\
        Thread & 212 & 0.047 & 215 & 15.0 \\
    \end{tabular}
\end{table}

For the subsequent analysis, 100 measurement sequences are recorded for each MMR, which are kept fixed at the center of the measuring coils with $\boldsymbol{p}_\mathrm{rotor}\approx{\vec{o}}$ in world coordinates. The measured data $\boldsymbol{u}(t,\, {\vec{o}},\, \boldsymbol{R})$ can be specified in the unit of the magnetic moment derivative $\left(\left[\frac{\mathrm{d}m}{\mathrm{d}t}\right]=\frac{\mathrm{Am}^2}{\mathrm{s}}\right)$ with the help of a system calibration $\boldsymbol{P}^{-1}$ according to eq.~\eqref{eq:u(t,p,R)} via
\begin{equation}
    \left(\frac{\mathrm{d}\boldsymbol{m}(t,\, \boldsymbol{R})}{\mathrm{d}t}\right)^\mathrm{meas.}
    = -\boldsymbol{P}^{-1} \ \boldsymbol{u}(t,\, {\vec{o}},\, \boldsymbol{R}) \; ,
\end{equation}
where
\begin{equation}
    \boldsymbol{P}^{-1} := \boldsymbol{P}^{-1}(\,\boldsymbol{p}_\mathrm{rotor}\,) \; \forall \; \boldsymbol{p}_\mathrm{rotor} \approx {\vec{o}}
\end{equation}
which is approximately constant for all $\boldsymbol{p}_\mathrm{rotor}$ that are in close proximity to the center of the Helmholtz coil arrangement. This is due to the reciprocity theorem and the nearly homogeneous field characteristic inside the Helmholtz coils.

\subsection{Analysis of Measurement Data}
\label{ssec:analysis_of_measurement_data}

The estimation algorithm is the key component in the evaluation pipeline as it derives the quantities for tracking and sensing from the measured data. \citeauthor{reiss_parameter_2026} compared different estimation algorithms and demonstrated that fitting a suitable model to the measured time signal yields the most precise results at the cost of longer calculation times \cite{reiss_parameter_2026}. It can also fulfill the secondary goal of validating the presented Jewel-MMR model in eq.~\eqref{eq:Jewel_ODE_norm}. Contrary to \citeauthor{reiss_parameter_2026}, this work does not fit the proposed MMR-models to the measured voltages, but rather to the calibrated signals in the magnetic moment derivative domain, assuming that
\begin{equation}
    \left(\frac{\mathrm{d}\boldsymbol{m}(t,\, \boldsymbol{R})}{\mathrm{d}t}\right)^\mathrm{meas.}
    \approx \left(\frac{\mathrm{d}\boldsymbol{m}(t,\, \mathcal{P})}{\mathrm{d}t}\right)^\mathrm{model} \; , \qquad t\in[0,\,T] \; ,
\end{equation}
where $\mathcal{P}$ is the set of all fitted parameters. This way, estimates of the MMR's orientation $\boldsymbol{R}$ and magnetic moment $m_\mathrm{r}$ can also be retrieved, which otherwise could not be extracted without the knowledge of the system calibration $\boldsymbol{P}^{-1}$.

The model trajectory $\left(\frac{\mathrm{d}\boldsymbol{m}(t,\, \mathcal{P})}{\mathrm{d}t}\right)^\mathrm{model}$ is obtained by numerically integrating the respective equation of motion, namely eq.~\eqref{eq:Thread_ODE_norm} for the Thread-MMR and eq.~\eqref{eq:Jewel_ODE_norm} for the Jewel-MMR, in combination with eq. \eqref{eq:projection}, which gives the equivalent magnetic moment derivative in world coordinates. The initial conditions for the ODEs are defined analogously to \citeauthor{reiss_parameter_2026} by
\begin{equation}
    \varphi(0) = \varphi_\mathrm{max} \cos\psi_\mathrm{start}
\end{equation}
and
\begin{equation}
    \dot\varphi(0) = -\dot\varphi_\mathrm{max} \sin\psi_\mathrm{start} \;.
\end{equation}
They are parameterized with the maximum deflection angle $\varphi_\mathrm{max}$ and phase $\psi_\mathrm{start}$, which is the oscillation phase at the beginning of the receive window. The maximum angular velocity follows from energy conservation as
\begin{equation}
    \dot\varphi_\mathrm{max} = \omega_\mathrm{nat} \sqrt{2 - 2\cos\varphi_\mathrm{max}} \; .\text{~\cite{reiss_parameter_2026}}
    \label{eq:dotvarphimax}
\end{equation}

The search for the 9 dependent parameters of the rotation matrix $\boldsymbol{R}$ is also simplified through a parameterization, here via Rodrigues' rotation formula
\begin{equation}
    \boldsymbol{R} = \cos\theta \, \boldsymbol{I} + \left(1-\cos\theta\right)\boldsymbol{r}\boldsymbol{r}^\mathsf{T}+\sin\theta
        \begin{pmatrix}
               0 & -r_z &  r_y \\
             r_z &    0 & -r_x \\
            -r_y &  r_x &    0 \\
        \end{pmatrix} ,
\end{equation}
with rotation angle and axis
\begin{equation}
    \theta = \left\vert\, \boldsymbol{\varrho}\, \right\vert \qquad \mathrm{and} \qquad \boldsymbol{r} = \frac{\boldsymbol{\varrho}}{\theta} \; ,
\end{equation}
where $\boldsymbol{\varrho} \in \mathbb{R}^3$ is the Rodrigues vector of just 3 independent parameters that can fully describe the MMR's orientation within the coil arrangement. The viscous damping enters the fit through the lumped parameter
\begin{equation}
    \lambda := \frac{\omega_\mathrm{nat}}{2Q} \;.
    \label{eq:lambda}
\end{equation}

The sign function within the friction term in the Jewel-MMR's ODE makes it a stiff and non-smooth problem, which is hard to solve numerically. To increase solver stability for the Jewel-MMR model, the sign function is approximated using the differentiable hyperbolic tangent function. Also, a different solver algorithm\footnote{\textit{Rodas5P}-algorithm from \href{https://docs.sciml.ai/DiffEqDocs/stable/}{\textit{DifferentialEquations.jl}} package \cite{rackauckas_differentialequations_2017,steinebach_construction_2023}} than in \cite{reiss_parameter_2026} is used, which is suited for stiff ODEs \cite{rackauckas_differentialequations_2017,steinebach_construction_2023}. A Levenberg-Marquardt algorithm\footnote{\textit{curve\_fit}-function from \href{https://juliapackages.com/p/lsqfit}{\textit{LsqFit.jl}} package} was used as the optimizer.

In addition to the estimation algorithm, the length of the fitted evaluation period $T$ also determines the precision of the extracted parameter. In practice, the ring-down is usually not recorded until the MMR has fully come to rest. On one hand, because the signal eventually disappears into the noise at small deflection angles, and on the other hand, because a higher measurement repetition rate is generally chosen to increase the temporal resolution of tracking and sensing. MMRs are therefore commonly operated close to the maximum stable deflection angle. To simulate this realistic usage scenario, the measured data is split into separate datasets. The datasets Jewel$_\mathrm{N50}$ and Thread$_\mathrm{N50}$ contain only the first approximately $T=0.24\,\mathrm{s}$ of data, corresponding to roughly 50 oscillations at the reference frequency of $212\,\mathrm{Hz}$. The datasets Jewel$_\mathrm{D30}$ and Thread$_\mathrm{D30}$ cover a larger portion of the ring-down, namely all samples up to the point at which the MMR signal in the channel with the largest response has decayed to 30\% of its maximum value. This is equivalent to a 70\% drop in the $x$-channel, since the MMR's $v$-axis is chosen to approximately align with the global $x$-axis in the measurement setup. Since the estimation algorithm has a larger number of samples available from these D30 datasets, it is assumed to yield more reliable results. Following this, the best fit obtained from each D30 dataset additionally serves as a substitute for the ground truth regarding the orientation error calculation. For 100 recorded sequences per MMR, the standard deviation of the estimated natural frequency and the mean angular error of the estimated orientation relative to the D30 reference are used as surrogate metrics for the achievable sensing and tracking performance, respectively. As the reference is taken per MMR from its own measurements, the angular deviation quantifies the repeatability of the orientation estimate rather than an absolute orientation accuracy.

Preliminary results of the model fit for the Jewel-MMR showed that, for short evaluation periods $T$ where the signal amplitude barely decays, the two damping terms $\lambda$ and $\frac{\hat{T}_\mathrm{f}}{I}$ cannot be separated from one another. The linearized parameter covariance obtained from the Levenberg-Marquardt Jacobian quantifies this ambiguity. $\lambda$ and $\frac{\hat{T}_\mathrm{f}}{I}$ are correlated with a coefficient of $-0.9998$ for the Jewel$_{\mathrm{N50}}$ dataset. They introduce an additional degree of freedom that prevents unambiguous convergence of the fit. For the Jewel$_{\mathrm{N50}}$ dataset, $\lambda$ is therefore fixed to the value extracted from the model fit results of the Thread$_\mathrm{D30}$ dataset, since both MMRs are assumed to be subject to similar viscous damping. This neglects the effect of the thread but, as will be shown, is a more robust estimate than the $\lambda$ obtained from the Jewel$_\mathrm{D30}$ dataset, where the longer observation window relaxes the correlation between the damping terms to at least $-0.97$.

Finally, the resulting parameter sets for each dataset are summarized in Table~\ref{tab:model_fit_params}. They contain the desired parameters that are to be used to quantify the performance of both MMR variants. For the secondary goal of verifying the proposed Jewel-MMR model, the coefficient of determination $R^2$ is to be used, which is derived directly from the residual vector of the solver's minimization function. 

\begin{table}[pos=h]
    \renewcommand{\arraystretch}{1.3}
    \centering
    \caption{Model Fit Parametersets}
    \label{tab:model_fit_params}
    \begin{tabular}{l|l|c}
        Dataset & Parameterset $\mathcal{P}$ & DoF \\[1pt]
        \hline
        $\mathrm{Jewel}_{\mathrm{N}50}$ & $\left\{ \omega_\mathrm{nat}, \; \boldsymbol{\varrho}, \; m_\mathrm{r}, \;\,\quad
        \textstyle{\frac{T_\mathrm{f}}{I}}, \; \varphi_\mathrm{max}, \; \psi_\mathrm{start} \right\}$ & 8 \\
        $\mathrm{Jewel}_{\mathrm{D}30}$ & $\left\{ \omega_\mathrm{nat}, \; \boldsymbol{\varrho}, \; m_\mathrm{r}, \; \lambda, \; \textstyle{\frac{T_\mathrm{f}}{I}}, \; \varphi_\mathrm{max}, \; \psi_\mathrm{start} \right\}$ & 9 \\
        $\mathrm{Thread}_{\mathrm{N}50}$ & $\,\left\{ \omega_\mathrm{nat}, \; \boldsymbol{\varrho}, \; m_\mathrm{r}, \; \lambda,  \qquad\! \varphi_\mathrm{max}, \; \psi_\mathrm{start} \right\}$ & 8 \\
        $\mathrm{Thread}_{\mathrm{D}30}$ & $\,\left\{ \omega_\mathrm{nat}, \; \boldsymbol{\varrho}, \; m_\mathrm{r}, \; \lambda,  \qquad\! \varphi_\mathrm{max}, \; \psi_\mathrm{start} \right\}$ & 8 \\
    \end{tabular}
\end{table}

Furthermore, a robustness analysis is added to highlight the advantages of stronger signals due to higher oscillation angles. System noise with an average of $0.030$~Am$^2$s$^{-1}$ per channel\footnote{$\sigma_{\dot m,\,x}\approx 14$~mAm$^2$s$^{-1}$, $\sigma_{\dot m,\,y}\approx 21$~mAm$^2$s$^{-1}$, $\sigma_{\dot m,\,z}\approx 55$~mAm$^2$s$^{-1}$} is increased artificially by adding digital white noise, which makes this a simulation-based robustness analysis rather than a measurement at a physically degraded signal-to-noise ratio. Both routes are closely related. To first order, a larger distance between resonator and receive coils scales the coil sensitivity $\boldsymbol{P}(\boldsymbol{p}_\mathrm{rotor})$ in eq.~\eqref{eq:u(t,p,R)}, which lowers the signal at constant noise and is therefore equivalent to raising the noise at constant signal. Adding the noise digitally has the further advantage that both variants are exposed to the same noise statistics, so that the comparison is not confounded by misplacement or field inhomogeneities. Additionally, a mostly qualitative comparison of the manufacturing process is conducted.

\section{Results}
\label{sec:results}

\subsection{Manufacturability}
\label{ssec:manufacturability}
Starting with the evaluation of the manufacturing process, the times for the MMRs' manual manufacturing steps are listed in Table~\ref{tab:manufacturing_times}. The construction times are highly subjective, as they depend heavily on the individual assembling the sensors, and they describe the two prototypes built here rather than the designs in general. They serve here only as a rough guide to the complexity involved. For these prototypes, the Jewel-MMR takes in total less than half of the time to assemble with regard to the Thread-MMR, since the time-consuming step of precisely aligning and gluing the magnets needs to be performed only once instead of twice. Independently of who carries out the assembly, the two variants also differ in countable structural properties, listed in the lower part of Table~\ref{tab:manufacturing_times}. The Jewel-MMR requires one adhesive bond instead of three and one precision alignment step instead of two, and, as established in Section~\ref{ssec:types_of_manufacturing_errors}, it exposes a single independent source of tilt error instead of two. Both variants are assembled from six individual parts, so the advantage lies in the number of critical assembly operations rather than in the part count. The $f_\mathrm{nat}$-tuning process of the Thread-MMR is intentionally excluded from the evaluation, as it is merely a necessary step for the comparison to the Jewel-MMR.

\begin{table}[pos=h]
    \centering
    \caption{Manufacturing times and structural properties}
    \label{tab:manufacturing_times}
    \begin{tabular}{p{4cm}cc}
        \multirow{2}{*}{Manufacturing step}
            & \multicolumn{2}{c}{Time in min. for} \\
        \cmidrule(lr){2-3}
            & Jewel-/ & Thread-MMR \\
        \midrule
        Thread $\mapsto$ Interface cap        &  - &  6 \\
        Interface cap $\mapsto$ Rotor magnet  &  - &  6 \\
        Stator magnet $\mapsto$ Fixture       &  6 &  6 \\
        Final assembly of parts               &  5 &  5 \\
        \midrule
        Total production time in min.         & 11 & 23 \\
        \midrule
        \midrule
        Structural properties
            & \multicolumn{2}{c}{Count}\\
        \midrule
        Individual parts                      &  6 &  6 \\
        Adhesive bonds                        &  1 &  3 \\
        Precision alignment steps             &  1 &  2 \\
        Independent tilt-error sources        &  1 &  2 \\
    \end{tabular}
\end{table}

\subsection{Quality of the Model-Fit}
\label{ssec:quality_of_modelfit}

\begin{figure*}[pos=t]
    \centering
    \includegraphics[width=\textwidth]{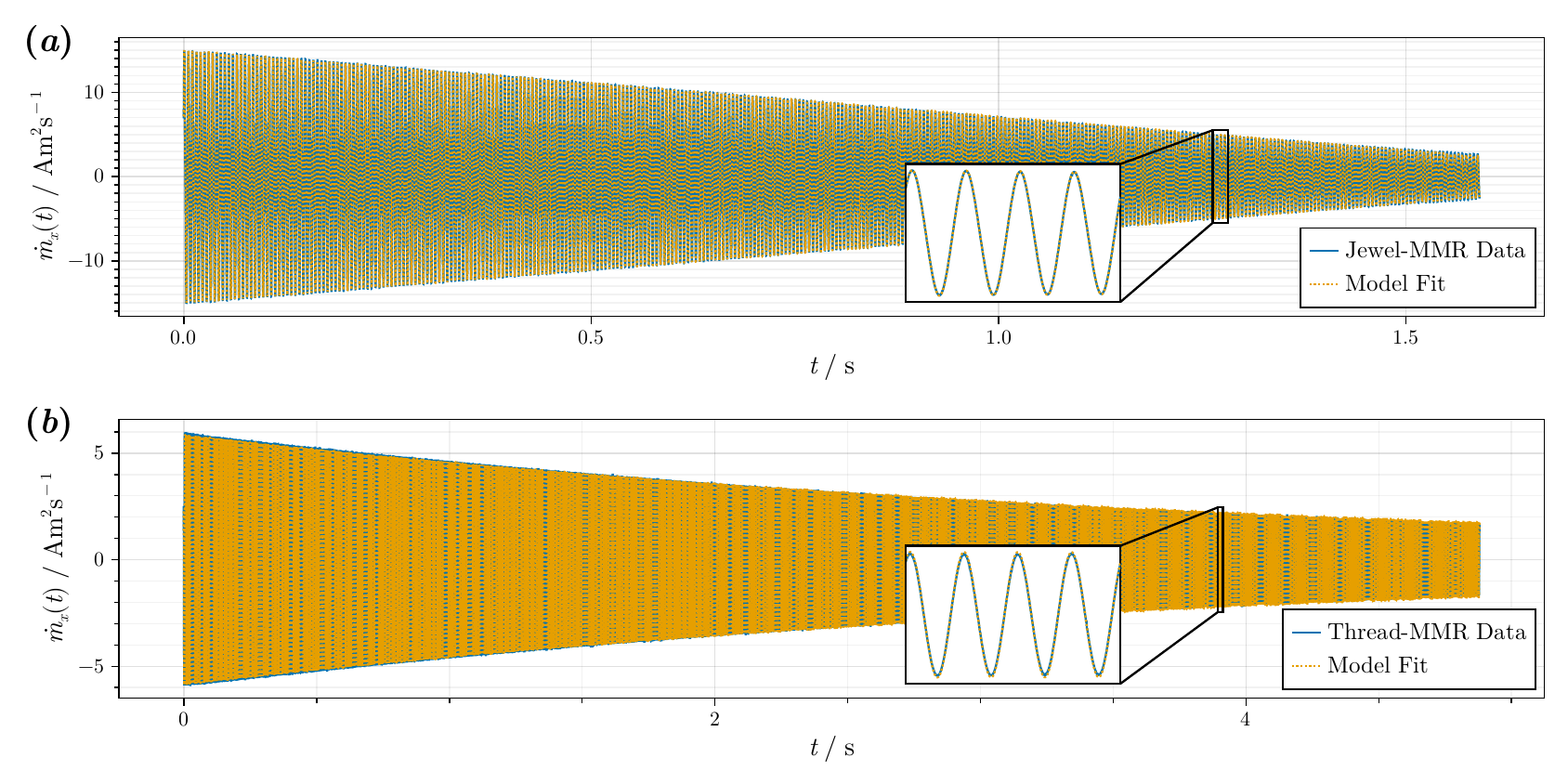}
    \plotspace
    \caption{Model fit results for \textit{(a)} the Jewel-MMR and \textit{(b)} the Thread-MMR. Both figures share the same grid spacing, but the axes are not scaled equally. The inlet plots show a zoomed view of roughly 4 oscillations each to highlight the visual agreement between the model and the measurement data.}
    \label{fig:fit_plot}
\end{figure*}

Figure \ref{fig:fit_plot} shows representative results for two model fits. The $x$-channel outputs of the fitted models are plotted against the corresponding measurement data from the D30 datasets for the Jewel- and Thread-MMR. Due to the chosen orientation, the $x$-channel output corresponds approximately to the $v$-projection of the MMR signals. Table~\ref{tab:fit_results_quality} quantifies the overall high visual agreement visible in the graph for the model fits across each dataset. It shows the median coefficient of determination $R^2$, the $\mathrm{nRMSE}$, that is, the RMSE normalized to the signal's RMS value in the specific dataset, and the median computation time. The N50 datasets each achieved higher levels of determination than their D30 counterparts while requiring less computation time. For the Thread-MMR, the increase in calculation time is only due to the larger observation window and therefore higher number of time samples that must be fitted. For the Jewel-MMR, an additional free parameter, as shown in Table~\ref{tab:model_fit_params}, is also a contributing factor. Regardless, the higher calculation times for the Jewel-MMR datasets in general are due to the stiff and non-smooth nature of the ODE in eq.~\eqref{eq:Jewel_ODE_norm}, which must be solved using solvers with higher accuracy and other algorithms as stated in Section \ref{ssec:analysis_of_measurement_data}.

\begin{table}[pos=ht]
    \renewcommand{\arraystretch}{1.3}
    \centering
    \caption{Overview of median fit results}
    \label{tab:fit_results_quality}
    \begin{tabular}{l|cc|c}
        Dataset & $R^2$ $/$ \% & $\mathrm{nRMSE}$ $/$ \% & $\mathrm{median}\left(t_\mathrm{comp.}\right)$ \\[1pt]
        \hline
        $\mathrm{Jewel}_{\mathrm{N}50}$ & 99.98 $_{\mathrm{min~} 99.76}$ & 1.38 & 2 min 9 s \\
        $\mathrm{Jewel}_{\mathrm{D}30}$ & 99.81 $_{\mathrm{min~} 95.53}$ & 4.39 & 22 min 26 s \\
        $\mathrm{Thread}_{\mathrm{N}50}$ & 99.98 $_{\mathrm{min~} 99.97}$ & 1.42 & 3 s \\
        $\mathrm{Thread}_{\mathrm{D}30}$ & 99.89 $_{\mathrm{min~} 99.88}$ & 3.33 & 49 s \\
    \end{tabular}
\end{table}

\subsection{Primary Results}
\label{ssec:primary_results}

Table~\ref{tab:model_params_fit_result} shows the mean and empirical standard deviation of the model parameters from Table~\ref{tab:model_fit_params} over the 100 recorded sequences, which are the primary results of the model fit for the different datasets. The reported standard deviations are the empirical parameter uncertainty of the estimates and are used as such throughout. The only exception is the rotation vector $\boldsymbol{\varrho}$, which is specified in terms of deviation or error angle relative to a reference vector $\boldsymbol{\varrho}_\mathrm{ref}$ based on the best estimate for the respective D30 dataset. For each MMR this is defined as the model fit with the highest $R^2$-value within the D30 dataset. The deviations of the estimates for the frequency and this error angle characterize the precision with which the sensing and tracking quantities are recovered from the respective MMR. Since $\boldsymbol{\varrho}_\mathrm{ref}$ is taken from the measurements themselves rather than from an independent optical or mechanical reference, the error angle quantifies the repeatability of the orientation estimate and not its absolute accuracy. Here the parameter estimation of the Thread-MMR achieves higher precision in both domains. Similar results emerge for the remaining model parameters divided into sensor and measurement characteristics as of Table~\ref{tab:model_params_fit_result}. Generally, the increased number of samples from the N50 to the D30 datasets seems to have a negligible effect on the sensing precision per MMR, since the standard deviations are similar, but for the tracking precision, they can lower the mean orientation error significantly with a significance level of $\alpha=0.001$.

\begin{table*}[pos=ht]
    \renewcommand{\arraystretch}{1.3}
    \centering
    \caption{Fitted model parameters}
    \label{tab:model_params_fit_result}
    \begin{tabular}{l|cc|ccc|cc}
        \multirow{2}{*}{Dataset} & \multicolumn{2}{c|}{Sensing and Tracking} &
            \multicolumn{3}{c|}{Sensor Characteristics} &
            \multicolumn{2}{c}{Meas. Characteristics} \\[0.1cm]
        & $f_{\mathrm{nat}} \; / \; \mathrm{Hz}$ &
            $\measuredangle (\boldsymbol{\varrho}, \boldsymbol{\varrho}_\mathrm{ref}) \; / \; {}^\circ$ &
            $m_\mathrm{r} \; / \; \mathrm{mAm}^2$ & $\lambda \; / \; \mathrm{s}^{-1}$ &
            $\frac{\hat{T}_\mathrm{f}}{I} \; / \; \mathrm{rad\,s}^{-2}$ & $\varphi_\mathrm{max}\; / \; {}^\circ$ &
            $\psi_\mathrm{start} \; / \; {}^\circ$ \\[0.1cm]
        \hline
        $\mathrm{Jewel}_{\mathrm{N}50}$ & 212.93 $\pm$ 0.13 & 0.132 $\pm$ 0.084 & 12.262 $\pm$ 0.014 & 0.2465 & 1063 $\pm$ 377 & 52.4 $\pm$ 0.5 & 219.1 $\pm$ 2.4 \\
        $\mathrm{Jewel}_{\mathrm{D}30}$ & 213.25 $\pm$ 0.19 & 0.079 $\pm$ 0.055 & 12.028 $\pm$ 0.143 & 0.0978 $\pm$ 0.1211 & 1324 $\pm$ 353 & 53.0 $\pm$ 0.6 & 219.2 $\pm$ 4.9 \\
        $\mathrm{Thread}_{\mathrm{N}50}$ & 215.53 $\pm$ 0.08 & 0.012 $\pm$ 0.005 & 11.647 $\pm$ 0.002 & 0.2466 $\pm$ 0.0012 & - & 21.8 $\pm$ 0.4 & 333.7 $\pm$ 1.2 \\
        $\mathrm{Thread}_{\mathrm{D}30}$ & 215.69 $\pm$ 0.08 & 0.004 $\pm$ 0.002 & 11.014 $\pm$ 0.004 & 0.2465 $\pm$ 0.0010 & - & 22.8 $\pm$ 0.4 & 335.7 $\pm$ 1.2 \\
    \end{tabular}
\end{table*}

Differences in mean values are also significant across datasets of the same MMR with $\alpha=0.001$ for all parameters except $\psi_\mathrm{start}$ for the Jewel-MMR and $\lambda$. Since they should actually be equal, it directly follows that, despite the high values for $R^2$ in Table~\ref{tab:fit_results_quality}, the model fit does not provide unbiased estimates and additional systematic effects are present which are not modeled by the underlying equations. Additionally, it is noticeable that the damping and friction terms for the Jewel-MMR datasets are subject to large fluctuations. In particular, $\lambda$ could not be reliably estimated, resulting in an empirically determined variance that violates the condition $\lambda \geq 0$. This does not compromise the results, since the damping contribution of $\lambda$ is negligibly small for the Jewel-MMR compared to $\frac{\hat{T}_\mathrm{f}}{I}$. As for the high fluctuations of $\frac{\hat{T}_\mathrm{f}}{I}$ itself, the dynamic evaluation periods $T_\mathrm{D30}$ of the Jewel$_\mathrm{D30}$ dataset also vary largely. The Jewel-MMR's $T_\mathrm{D30}$-values are around 1.28~$\pm$~0.31~s, whereas those of the Thread-MMR are at 4.90~$\pm$~0.01~s. This implies that the signal decay and therefore the friction torque actually varies and $\frac{\hat{T}_\mathrm{f}}{I}$ is not a constant sensor characteristic, as assumed.

\begin{table}[pos=ht]
    \renewcommand{\arraystretch}{1.3}
    \centering
    \caption{Approximate Quality Factors}
    \label{tab:quality_factors}
    \begin{tabular}{l|ccc}
        MMR & $\tilde{Q}_\mathrm{N50}$ & $\tilde{Q}_\mathrm{D30} $ & $\tilde{Q}_{T\to\infty}$ \\[1pt]
        \hline
        Jewel & $\approx1327$ & $\approx897$ & $\approx641$ \\
        \hline\hline
        Thread & \multicolumn{3}{c}{$Q\approx2750$} \\
    \end{tabular}
\end{table}

\begin{figure*}[pos=ht]
    \centering
    \includegraphics[width=\textwidth]{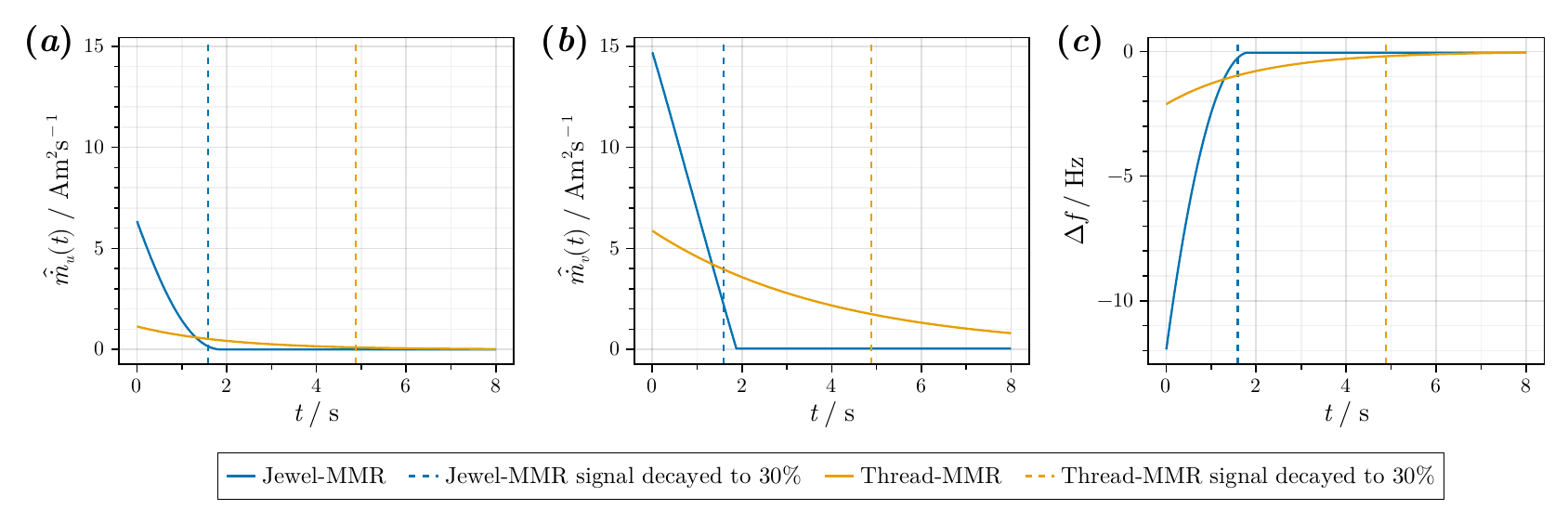}
    \plotspace
    \caption{Amplitude plots for \textit{(a)} $u$-channel and \textit{(b)} $v$-channel signal projection. Subplot \textit{(c)} depicts the frequency difference~$\Delta f$ to the natural frequency $f_\mathrm{nat}$ for each MMR. The dashed vertical lines indicate the point in time, where the signal in the $x$-Channel of the measurement system decayed to 30\% of its initial value.}
    \label{fig:amplitude_plot}
\end{figure*}

For reference, Table~\ref{tab:quality_factors} lists the equivalent quality factors according to eq.~\eqref{eq:tilde-Q} for the Jewel-MMR across different time periods $T$. They range from 1327 for the Jewel$_\mathrm{N50}$ dataset with $T\approx0.24$~s down to 641 for $T\to\infty$ and are roughly two to four times smaller than the Thread-MMR's Q-factor with $Q\approx2750$. To provide a comprehensive overview of the performance of both MMRs, despite the limitations of the equivalent Q-factor, Fig.~\ref{fig:amplitude_plot} shows the envelopes $\hat{\dot m}(t)$ of the projected time signals and frequency traces from both sensors side-by-side. They are derived from the fitted models and refer to the MMRs' local $uvw$ coordinate system. It demonstrates that, while the signal response of the Jewel-MMR decays significantly faster, it also starts with a higher amplitude. This effect can be attributed to the initially larger deflection angle of $\varphi_\mathrm{max}\approx53^\circ$ for the Jewel-MMR versus only $\varphi_\mathrm{max}\approx23^\circ$ for the Thread-MMR according to the results in Table~\ref{tab:model_params_fit_result}. Since the instantaneous frequency of the MMRs is directly related to the deflection angle, this in turn means that the plotted frequency change for the Jewel-MMR is also noticeably steeper and exhibits a larger range (see Fig. \ref{fig:amplitude_plot}\textit{c}).

\begin{figure*}[pos=ht]
    \centering
    \includegraphics[width=\textwidth]{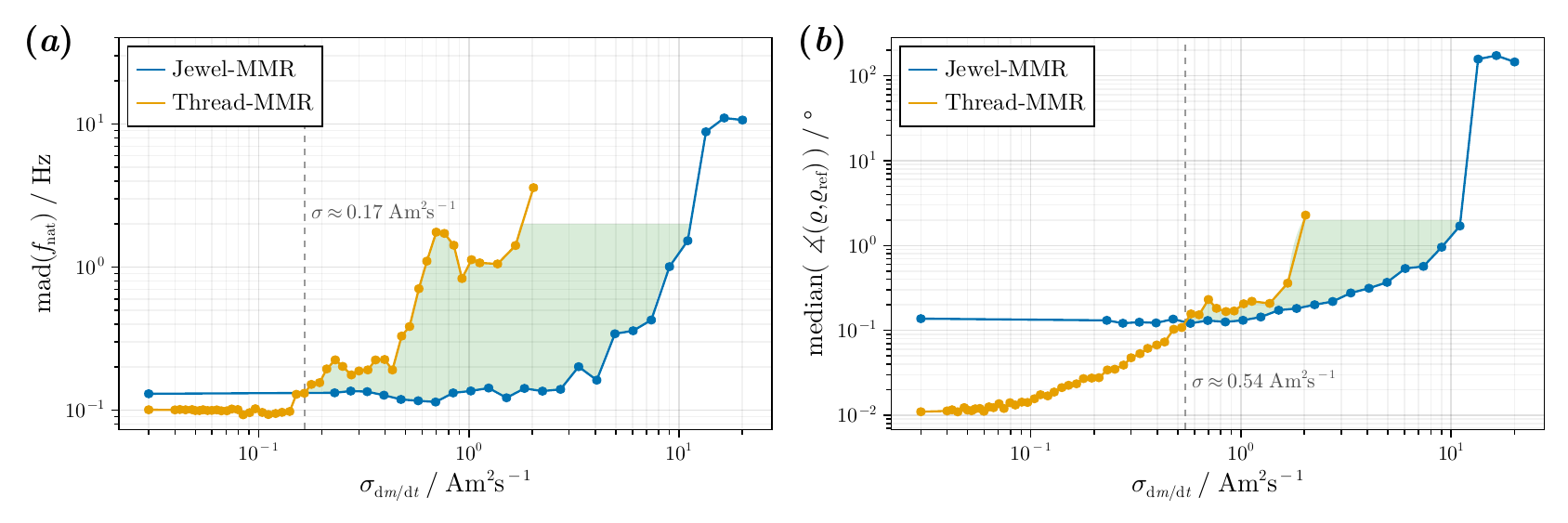}
    \plotspace
    \caption{Median absolute deviation of the natural frequency estimates in \textit{(a)} and median orientation error in \textit{(b)} over the digitally increased system noise level for the N50 datasets of both MMR variants. The dashed vertical lines mark the break-even points at which the estimation based on the Jewel-MMR becomes more precise than that of the Thread-MMR.}
    \label{fig:noise_plot}
\end{figure*}

Another effect of larger deflection angles is higher signal amplitudes, which make the estimation of both quantities more robust against noise. To highlight this effect, Fig. \ref{fig:noise_plot} plots the median deviation of the natural frequency estimate and the median of the error angle over the noise level for the N50 datasets. Using quality metrics based on the median makes the analysis more robust against individual model fits that fail due to the added noise. The dashed lines in the plots visualize the break-even points for the noise levels at which the fitting of Jewel-MMR becomes more precise than the one of the Thread-MMR. For a sensing application, this is at a noise level of approximately $0.17$~Am$^2$s$^{-1}$, for tracking at around $0.54$~Am$^2$s$^{-1}$.

\section{Discussion}
\label{sec:discussion}

Parameters $f_\mathrm{nat}$ and $\frac{\hat{T}_\mathrm{f}}{I}$, extracted from the model, lie within a $\pm20\%$ tolerance window regarding the values of the design equations evaluated in eqs.~\eqref{eq:fnat} and~\eqref{eq:TfoverI}, respectively. For $m_\mathrm{r}$, the datasheet specifies a tolerance of $\pm 12\%$, whereas the measured values consistently fall below the design equation \eqref{eq:m_r}, by 16 to 25\%. Magnets can lose part of their magnetization due to various factors such as elevated temperatures, mechanical shocks, or aging over time. The observed, one-sided deviations are attributed to such effects rather than to a systematic model error. Overall, the deviations remain within a reasonable range, validating the design equations as a useful tool when dimensioning MMRs and especially when considering different contact materials for future Jewel-MMRs.

While searching for the measurement parameters, it became apparent that the Jewel-MMR can be deflected to a much greater angle before side mode oscillations begin to affect the signal. The Jewel-MMR, thus, achieved a maximum deflection angle $\varphi_\mathrm{max}$ of around $53^\circ$ in the experiments, while the Thread-MMR reached only $23^\circ$. This resulted in an approximately five times stronger $u$-signal and three times stronger $v$-signal for the Jewel-MMR at the start of the receive windows. Due to that, in the simulation-based robustness analysis the estimates obtained from the Jewel-MMR become the more precise ones above a noise level of 0.17~Am$^2$s$^{-1}$ for the natural frequency and 0.54~Am$^2$s$^{-1}$ for the orientation. Section~\ref{ssec:manufacturing_errors_side_modes} proposed a connection between thread length or cup radius and side mode resilience in combination with manufacturing defects. Based on these considerations, the thread used in the experiments would be expected to show a slightly increased susceptibility to side modes at equal stator tilt defects when compared to the Jewel-MMR. However, all referenced publications in this work featuring Thread-MMR, some of which use shorter threads, showed similar values for $\varphi_\mathrm{max}$, when considering that the MMRs were excited with only 80\% of the found maximum excitation field strength. This indicates that a longer thread is unlikely to be the sole cause. One possible explanation for the reduced side mode susceptibility in the Jewel-MMR might be the point contact between the rotor and the jewel. This contact point could restrict or damp the lateral motion of the rotor to a higher degree compared to a Thread-MMR. Neither explanation can be separated from the other with the present data, since the side modes themselves were not measured. A direct test would include a series of otherwise identical MMRs built with stator fixtures tilted by defined amounts with an uncertainty of well below $1^\circ$ which lies beyond the scope of this work. However, the observations indicate how strongly the structural design of the bearing impacts the usable deflection range and thus the achievable SNR.

The manual production of MMRs results in a wide variation in quality. The Thread-MMR used in the experiments with a Q-factor of 2750 lies at the upper end of the spectrum according to \citeauthor{gleich_miniature_2023} \cite{gleich_miniature_2023} making it a strong candidate for comparison. The Jewel-MMR with an equivalent Q-factor of 1327 for the time period of the first 50 oscillations lies within the middle range of typical Thread-MMR Q-values. The Thread-MMR assembly requires more adhesive bonds and more precision alignment steps and is therefore more complex to manufacture. If the proposed link between manufacturing defects and side mode excitation holds, Thread-MMRs are likely subject to greater variance in quality.

The choice of evaluation method has a strong influence on both the achievable precision and the required computation time, and the two cannot be optimized independently. The model-fit estimator used here yields the most precise frequency and orientation estimates \cite{reiss_parameter_2026}, but for the Jewel-MMR, the higher model complexity, i.e. the stiff and non-smooth dry friction term, translates into substantially longer computation times than for the Thread-MMR. Frequency-domain estimators, in contrast, have a fixed computational cost and enable real-time evaluation, but their precision degrades as the frequency drift within the evaluation window increases \cite{reiss_parameter_2026}. Since the Jewel-MMR exhibits a markedly steeper frequency trace (Fig.~\ref{fig:amplitude_plot}\textit{c}), such estimators are inherently disadvantaged for this variant. Reducing the friction torque, and thereby the frequency drift, is therefore not only beneficial for the quality factor but a prerequisite for fast, real-time-capable evaluation with frequency-domain estimators.

It should be emphasized that the solver was not optimized for speed. The reported computation times can be reduced considerably for known use cases, for example, by fixing the sensor characteristics that are constant for a given specimen; namely $\lambda$ and $m_\mathrm{r}$ for both MMRs, as well as $\frac{T_\mathrm{f}}{I}$ for the Jewel-MMR. Another approach could be sub-sampling the measured data to obtain a fast initial estimate that is subsequently refined. The values reported in Section~\ref{sec:results} should therefore be read as an upper bound rather than a fundamental limit.

A further point concerns the ambiguity of the deflection angle, which is a key problem for sensing applications as the oscillation frequency depends strongly on the maximum deflection angle $\varphi_\mathrm{max}$. On one hand, a large deflection angle is desirable, as it increases the signal amplitude and allows $\varphi_\mathrm{max}$ to be recovered based on the ratio of the first and second harmonics, which correspond to the $v$- and $u$-oscillation amplitudes and are readily obtained from a frequency analysis. On the other hand, one may deliberately restrict operation to the approximately linear small-angle regime where the frequency becomes amplitude-independent. This, however, requires larger magnets to maintain the same signal-to-noise ratio.

Finally, although the model fits achieve very high coefficients of determination ($R^2 > 99.8\,\%$, Table~\ref{tab:fit_results_quality}), the estimated parameters differ significantly between the N50 and D30 datasets of the same MMR variant. This indicates that the estimates are biased and that systematic effects remain unmodeled. The most likely explanation is that quantities assumed constant in the model are in fact deflection-angle dependent. This applies in particular to the friction and damping terms, as already suggested by the $\cos^{4/3}\varphi$ behavior derived in Section~\ref{ssec:characteristics-of-jewel-bearings}. Larger fluctuations of $\lambda$ in the Jewel$_\mathrm{D30}$ suggest that even longer evaluation periods are unable to eliminate a certain ambiguity in the model with regard to the dominating $\frac{\hat{T}_\mathrm{f}}{I}$-term. As discussed for the Jewel$_\mathrm{N50}$, the degrees of freedom were reduced by fixing $\lambda$ to the expected value and simultaneously stabilizing $\frac{\hat{T}_\mathrm{f}}{I}$. For the Jewel$_\mathrm{D30}$ dataset, fixing $\lambda$ showed little change for $\frac{\hat{T}_\mathrm{f}}{I}$ but also a slightly reduced $R^2$-value as an additional degree of freedom can also compensate for unmodeled effects. For the model fit evaluation, the parameters $m_\mathrm{r}$, $\lambda$, and $\frac{\hat{T}_\mathrm{f}}{I}$ describing the sensors' characteristics were assumed to be constant in time. However, a closer examination of the measured time series for the Jewel-MMR revealed an actual fluctuation of the $\hat{T}_\mathrm{f}$ parameter. The most plausible mechanism is the contact point itself. The rotor of a Jewel-MMR is not fixed to anything, so between measurement sequences it is free to settle on a different part of its surface, and the contact then samples a different patch of the local roughness of the coating. Under this reading, $\hat{T}_\mathrm{f}$ is not a device constant but a quantity that is realized anew for each sequence, which would account for the observed scatter of $1063\pm377$~rad\,s$^{-2}$ in Table~\ref{tab:model_params_fit_result}. The limiting quantity is then not the form of the friction model but the surface finish of the commercial NiCuNiCr coating. Wear of the involved materials and an accumulation of dust within the MMR's housing act in a similar way and cannot be excluded. Their magnitude, however, is not predictable from the present data and only the latter could be removed by hermetically sealing the housing under cleanroom conditions. Ultimately, the proposed Jewel-MMR model captures the dominant dynamics but does not yet fully describe the resonator, which is also true for the established Thread-MMR model.

Lastly, several limitations delineate the scope of these findings. Most importantly, the presented experiments are not sensing or tracking experiments. Both resonators were measured at a single fixed pose without an applied measurand, so that only the precision of the underlying quantities is assessed. The step from fixed-pose precision to volumetric tracking has already been demonstrated for Thread-MMRs \cite{gleich_miniature_2023, reiss_parameter_2026}, and since the Jewel-MMR generates its signal by the same mechanism, that step is not repeated here. Likewise, the elevated noise levels were obtained by adding white noise to the recorded data rather than by an actual degradation of the signal-to-noise ratio, for instance through a larger distance between sensor and receive coils. As set out in Section~\ref{ssec:primary_results}, this was a deliberate choice, since it exposes both variants to the same noise statistics. The link between manufacturing defects and side mode excitation was established theoretically through the equivalent perpendicular force analysis, but it was not validated experimentally. Only the main mode behavior of the two variants was compared, while the side mode characteristics were not measured directly. Likewise, the dry friction model relies on a simplified, quasi-static treatment of the Hertzian contact and neglects effects such as wear, surface characteristics, and impurities. Furthermore, the specified manufacturing times are subjective and can only be used as one indicator for complexity of assembly. The structural properties behind them, namely the self-centering rotor and the absence of a second adhesive alignment step, are properties of the design rather than of hand assembly and act in the same direction under automated production. A quantitative transfer, however, would require yield statistics over a manufactured series and is not provided here. Within these constraints, the results nonetheless support the conclusions drawn below.

\section{Conclusion}
\label{sec:conclusion}

This work investigated the impact of the structural design on the performance of magneto-mechanical resonators, using a novel jewel bearing variant as a case study. To this end, the Jewel-MMR was introduced as an alternative bearing concept and compared against the established Thread-MMR at equal magnet size. A dynamic model including the angle-dependent dry friction torque was derived and experimentally supported, and a two-tiered evaluation framework was proposed to enable a fair comparison across both the operational and the manufacturing life-cycle of the sensors. From the model, a geometric trade-off between friction torque and predicted side mode susceptibility was identified as the central structural design decision for the jewel cup.

Generally, the Jewel-MMR exhibits a lower oscillation quality than the Thread-MMR, reflected in a faster signal decay and a smaller equivalent quality factor. It is nonetheless sufficient to recover both underlying quantities with a precision of the same order as that of the Thread-MMR. This is especially true for environments with elevated noise levels, since the Jewel-MMR could be excited to larger deflection angles, yielding a stronger signal or higher SNR. The larger stable deflection angle observed for the Jewel-MMR is consistent with an improved resilience against side modes, although the side modes themselves were not measured. If the measurement system is able to generate sufficiently strong magnetic fields, the Jewel-MMR matches or surpasses the thread variant in the precision of both the frequency and the orientation estimate once the noise level exceeds the break-even points reported by the simulation-based robustness analysis. The self-aligning property of the Jewel-MMR's rotor magnet reduces the number of potential sources of manufacturing defects, which is expected to benefit the oscillation stability. At the same time, it needs fewer adhesive bonds and fewer precision alignment steps, it was assembled in less than half the time for the prototypes built here, and it carries no thread whose failure would irreversibly disable the sensor. These properties are expected to gain relevance at smaller scales and in series production, where thread handling and adhesive bonding reach their practical limits, and in real-world application scenarios, where noise is typically more prominent than under laboratory conditions.

Future work should address the points left open here. The predicted relationship between manufacturing defects and side mode excitation calls for direct experimental validation, and the unmodeled deflection-angle dependence of the friction and damping terms motivates a refined contact model. With regard to the scattering of $\hat{T}_\mathrm{f}$ itself, the surface finish of the coating rather than the friction model form appears to be the limiting quantity. Proving this claim would require measuring the roughness before and after an endurance run while estimating the distribution of $\hat{T}_\mathrm{f}$ periodically. Since the Jewel-MMR generates its signal in the same way as the Thread-MMR, its suitability for sensing and tracking follows from the precision of the underlying quantities rather than from a direct demonstration. Dedicated experiments with an applied measurand and with a position estimate over an extended working volume are therefore the immediate next step. For sensing applications in particular, possible designs of a variable magnet distance in Jewel-MMRs have to be considered, since the natural frequency depends strongly on that distance \cite{faltinath_natural_2025, knopp_empirical_2024}. To reduce complexity, a variable stator position that is dependent on the measuring quantity instead of the rotor position, as is usually done for the Thread-MMR, may be advantageous. A compliant intermediate element between the stator mount and the housing would translate pressure into a change of magnet distance, as already demonstrated for Thread-MMRs \cite{merbach_wireless_2025}. A thermally expanding spacer would do the same for temperature, and a swellable functional layer beneath the stator mount for chemical quantities. Moreover, the proposed evaluation framework may serve as a basis for comparing the growing number of MMR concepts on common, life-cycle-aware grounds.

\section*{Acknowledgements}

The authors would like to thank Bernhard Gleich and Jürgen Rahmer for their advice on the strategy for implementing the evaluation routine, and Lars Woinar for his technical support.

\appendix
\numberwithin{equation}{section}
\numberwithin{figure}{section}
\numberwithin{table}{section}

\section{Proof of size independence for Jewel-MMRs}
\label{sup:proof_size_independence}

To assess the influence of the overall size of a Jewel-MMR on its oscillation quality, a simplified small-angle model is used, since only the relative behavior under uniform scaling of all geometric parameters is of interest here. For an oscillation with angular amplitude $\varphi_0$, the quality factor defined as the ratio of average stored energy to energy loss per cycle evaluates to
\begin{equation}
    Q(\varphi_0) := 2\pi \frac{\text{average stored energy}}{\text{energy loss per cycle}}
    = 2\pi \frac{\frac{1}{2} \mu_T \varphi_0^2}{4 T_\mathrm{f}\varphi_0} \;,
\end{equation}
where $\mu_T = m_\mathrm{r} B_0$ is the magnetic torsion stiffness in the small-angle regime and $4 T_\mathrm{f} \varphi_0$ the dry friction loss per cycle. Under uniform scaling with the magnet radius $r$, i.e., $R \propto r$ and $d \propto r$, the magnetic moment scales as $m_\mathrm{r} \propto r^3$ while the stator field at the rotor position $B_0 \propto m_\mathrm{r}/d^3$ remains constant, hence $\mu_T \propto r^3$. According to eq.~\eqref{eq:fm(phi)}, the contact force scales as $F \propto r^6/d^4 \propto r^2$, the pressure circle radius by eq.~\eqref{eq:a} as $a \propto (R^* F)^{1/3} \propto r$, and the friction torque by eq.~\eqref{eq:tf} as $T_\mathrm{f} \propto a F \propto r^3$. Both $\mu_T$ and $T_\mathrm{f}$ therefore scale with $r^3$ and $Q(\varphi_0)$ is independent of the overall size.

\section{Detailed derivation of the friction torque of a jewel bearing}
\label{sup:derivation-friction-torque-jewel-bearing}

The equation of the friction torque for a jewel bearing, as presented in Section \ref{ssec:characteristics-of-jewel-bearings}, is derived from Hertz's theory in the field of contact mechanics \cite[see][chapter 4]{johnson_contact_1985}.
In Hertzian contact theory, two deformable bodies, here the jewel cup and the rotor magnet, are replaced by one equivalent body deforming against a rigid half-plane. For this simplification, an equivalent curvature $R^*$ and composite Young's modulus $E^*$ for the deformable body are defined through the following derivations.

\begin{figure}[pos=ht]
    \centering
    \includegraphics[width=0.7\columnwidth]{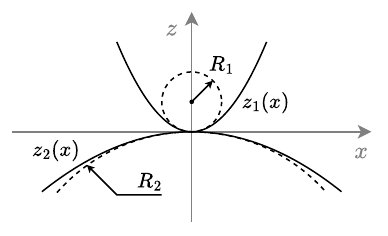}
    \vspace{-12pt}
    \caption{Definitions of the simplified Hertz contact theory for a rotationally symmetric case of two spherical bodies with $R_1$ and $R_2$ touching each other in the $xz$-plane. $z_1(x)$ and $z_2(x)$ are parabolas that approximate the spherical surfaces at $x\approx0$ and that are used to calculate an equivalent relative curvature.}
    \label{fig:hertz-contact}
\end{figure}

Figure~\ref{fig:hertz-contact} depicts the rotationally symmetric case with respect to the $z$-axis and two spheres with radii $R_1$ and $R_2$ touching each other in the $xy$-plane. Here, only the $xz$-plane is considered for the sake of symmetry. The equivalent radius $R^*$ is then derived via parabola approximations through Taylor's expansion. Let $z_1(x)$ be the height of the first surface and $z_2(x)$ the height of the second. Then, for small $x$ the following applies:
\begin{equation}
    z_1(x) \approx \frac{x^2}{2R_1}, \qquad z_2(x) \approx -\frac{x^2}{2R_2} \;.
\end{equation}
The signs of $z_1(x)$ and $z_2(x)$ are implicitly derived from the definition of the centers of the spheres, i. e. $R_1\boldsymbol{e}_z$ and $-R_2\boldsymbol{e}_z$. The relative profile height is defined as
\begin{equation}
    z^*(x)=z_1(x)-z_2(x) \approx \frac{x^2}{2} \left( \frac{1}{R_1} + \frac{1}{R_2} \right)
\end{equation}
which has the form of a parabola again
\begin{equation}
    z^*(x) \approx \frac{x^2}{2R^*}
\end{equation}
with the relative curvature
\begin{equation}
    R^* = \frac{1}{\frac{1}{R_1}+\frac{1}{R_2}} = \frac{R_1 R_2}{R_2+R_1}\;.
\end{equation}
Substituting $R_1$ with the radius $r$ of the rotor magnet and $-R_2$ with the jewel cup radius $R$, the geometric dependency introduced as the third term under the root in eq.~\eqref{eq:a} can be obtained as
\begin{equation}
\boxed{
    R^* = \frac{rR}{R-r}
}\;.
    \label{eq:Rstar}
\end{equation}

Similarly, for the composite Young's modulus $E^*$, if two elastic bodies are pushed together, the same pressure acts on both contact partners and their surfaces deform. The total deformation $\delta_\mathrm{total}$ as well as the total material compliance $C_\mathrm{total}$ are therefore the sum of their parts:
\begin{equation}
    \delta_\mathrm{total} = \delta_1 + \delta_2 \quad \mathrm{and} \quad
    C_\mathrm{total} = C_1 + C_2
\end{equation}
And with the material compliance for the Hertzian compression
\begin{equation}
    C_i = \frac{1-\nu_i^2}{E_i}
\end{equation}
follows
\begin{equation}
    C_\mathrm{total} := \frac{1}{E^*} = \frac{1-\nu_1^2}{E_1} + \frac{1-\nu_2^2}{E_2}\;.
\end{equation}
Using the common approximation for the Poisson's ratio in homogeneous materials, i.e. $\nu_1 \approx \nu_2 \approx \nu$, the inverse of $E^*$ can be written as
\begin{equation}
\boxed{
    \frac{1}{E^*}
    = \frac{1-\nu^2}{\frac{E_1 E_2}{E_1+E_2}}
}
    \label{eq:Estar}
\end{equation}
just like the second term under the root in eq.~\eqref{eq:a}.

For the following considerations on the friction torque $T_\mathrm{f}$ as of eq.~\eqref{eq:tf}, let the contact circle
\begin{equation}
    a = \sqrt[3]{\frac{3FR^*}{4E^*}}
    \label{eq:araw}
\end{equation}
and the pressure distribution
\begin{equation}
    p(\rho) = p_0 \sqrt{1 - \left( \frac{\rho}{a} \right)^2},
    \qquad 0\le \rho \le a
\end{equation}
for a spherical Hertz contact with the maximum pressure of
\begin{equation}
p_0=\frac{3F}{2\pi a^2},
\end{equation}
be given. They follow directly from Hertz's theory; a detailed derivation can be found in \cite[chapter 4.2]{johnson_contact_1985}. Then, the area of contact on a ring in the half-plane with radius $\rho$ has the shear stress $\tau_\mathrm{sh} = \mu p(\rho)$. The torque of the annular element (annulus) is
\begin{equation}
    \mathrm{d}T_\mathrm{f} =\underbrace{\mu \cdot p(\rho)}_{\text{shear}} \times
                        \underbrace{2\pi \rho \cdot \mathrm{d}\rho}_{\text{ring area}} \times
                        \underbrace{\rho}_{\text{lever}}\;,
\end{equation}
and therefore
\begin{equation}
    T_\mathrm{f} = \int_0^a \mathrm{d}T_\mathrm{f} = 2\pi\mu\int_0^a p(\rho)\, \rho^2\, \mathrm{d}\rho\;.
\end{equation}
Using the substitution $\rho=au$ with $u\in[0,1]$, the following applies
\begin{equation}
    T_\mathrm{f} = 2\pi\mu p_0 a^3 \int_0^1 u^2\sqrt{1-u^2} \mathrm{d}u \;.
\end{equation}
The integral can be calculated as
\begin{equation}
    \int_0^1 u^2\sqrt{1-u^2} \, \mathrm{d}u  = \frac{\pi}{16}\;,
\end{equation}
where
\begin{equation}
    T_\mathrm{f} = 2\pi\mu p_0 a^3\cdot\frac{\pi}{16} \;.
    \label{eq:T_fp_0}
\end{equation}
Finally, inserting $p_0=\frac{3F}{2\pi a^2}$ into eq. \eqref{eq:T_fp_0} yields
\begin{equation}
\boxed{T_\mathrm{f} = \frac{3}{16} \pi a\mu F}\;,
\end{equation}
which is eq.~\eqref{eq:tf}, introduced in Section \ref{ssec:characteristics-of-jewel-bearings} and with eqs. \eqref{eq:Rstar} and \eqref{eq:Estar} in eq. \eqref{eq:araw} follows for the special jewel bearing case considered in this work
\begin{equation}
    \boxed{
        a  = \sqrt[3]{  \frac{3}{4} \cdot
                        \frac{1-\nu^2}{\frac{E_1 E_2}{E_1+E_2}} \cdot
                        \frac{rR}{R-r} \cdot
                        F}
    }
\end{equation}
which is equal to eq. \eqref{eq:a}.

\section{Additional manufacturing defects}
\label{sup:additional_manufacturing_defects}

\begin{figure}[pos=htbp]
    \centering
    \includegraphics[width=\columnwidth]{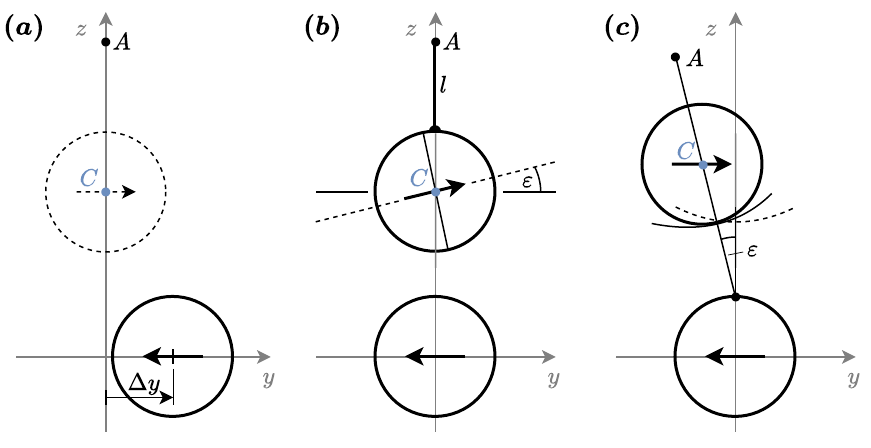}
    \caption{Schematic definition of the additional manufacturing defects, analogous to Fig.~\ref{fig:math_description_angle_defects}: \textit{(a)} lateral displacement of the stator by $\Delta y$ along its magnetization orientation, \textit{(b)}~tilt of the rotor (Thread-MMR), and \textit{(c)} tilt of the jewel (Jewel-MMR) by an error angle $\varepsilon$. As before, the new equilibrium position~$C$ of the rotor and the anchor point $A$ define the new axis of rotation, about which the rotor is deflected by angle $\varphi$.}
    \label{fig:math_description_delta_y-defect}
\end{figure}

Section \ref{ssec:types_of_manufacturing_errors} introduced all positioning and alignment defects of Thread- and Jewel-MMRs, of which only the tilted stator was analyzed in Section \ref{ssec:manufacturing_errors_side_modes} with respect to the side mode susceptibility of the MMRs. This appendix supplements these considerations with the remaining manufacturing defects, namely the tilted rotor and the lateral displacement for the Thread-MMR, as well as the tilted jewel and the lateral displacement for the Jewel-MMR. Figure~\ref{fig:math_description_delta_y-defect} depicts, analogously to Fig.~\ref{fig:math_description_angle_defects}, the mathematical definitions of the lateral displacement $\Delta y$ in sub-figure~\textit{(a)} for both MMR variants and of the respective error angle $\varepsilon$ for the rotor tilt in sub-figure~\textit{(b)} and for the jewel tilt in sub-figure~\textit{(c)}.

\begin{figure*}[pos=htbp]
    \centering
    \includegraphics[width=\textwidth]{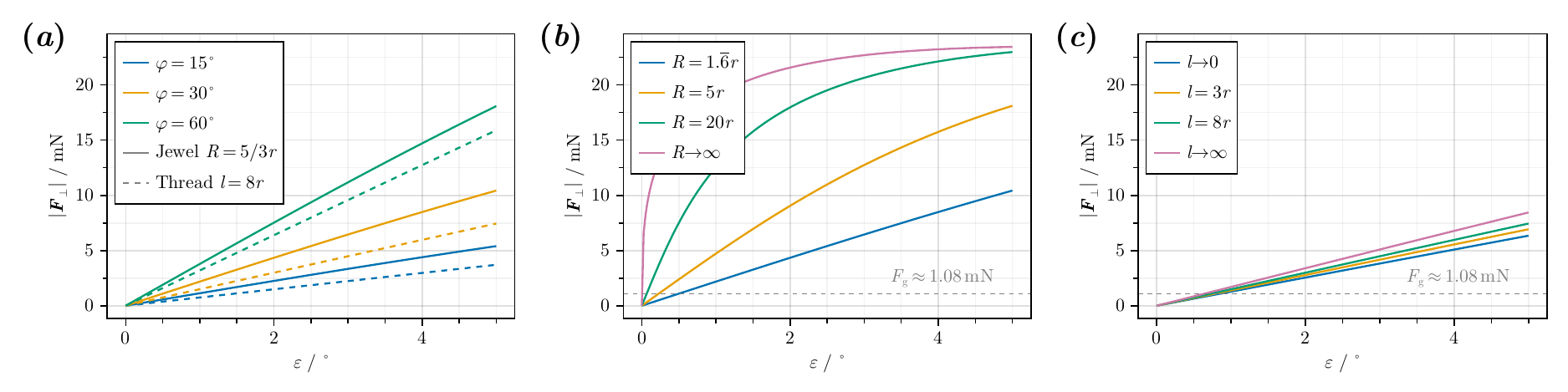}
    \plotspace
    \caption{Simulation results of the equivalent perpendicular force magnitude $\widetilde{F}_\perp$ with a tilted rotor in Thread-MMRs and a tilted jewel in Jewel-MMRs of error angle $\varepsilon$. Subplot \textit{(a)} shows multiple deflection angles $\varphi$ at the cup radius $R=\frac{5}{3}r$ for the Jewel-MMR and at the thread length $l=8r$ for the Thread-MMR. Subplots \textit{(b)} and \textit{(c)} show results with a fixed deflection angle $\varphi=30^\circ$ for multiple cup radii and thread lengths respectively. The magnets of both MMRs have a radius of $r=1.5$~mm, density of $\rho=7800$~kg~m$^{-3}$, remanent magnetization of $B_\mathrm{r}=1.3$~T and an initial distance of $d=4.7$~mm.}
    \label{fig:F_tan_over_rotor_epsilon}
\end{figure*}

\begin{figure*}[pos=htbp]
    \centering
    \includegraphics[width=\textwidth]{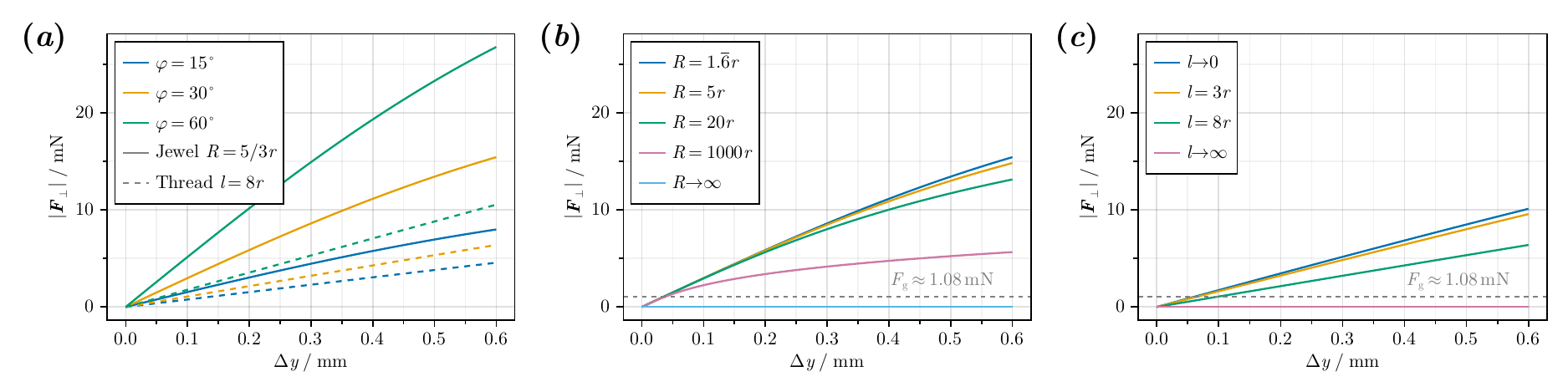}
    \plotspace
    \caption{Simulation results of the equivalent perpendicular force magnitude $\widetilde{F}_\perp$ in MMRs with manufacturing defects of lateral displacement $\Delta y$. Subplot \textit{(a)} shows multiple deflection angles $\varphi$ at the cup radius $R=\frac{5}{3}r$ for the Jewel-MMR and at the thread length $l=8r$ for the Thread-MMR. Subplots \textit{(b)} and \textit{(c)} show results with a fixed deflection angle $\varphi=30^\circ$ for multiple cup radii and thread lengths respectively. The magnets of both MMRs have a radius of $r=1.5$~mm, density of $\rho=7800$~kg~m$^{-3}$, remanent magnetization of $B_\mathrm{r}=1.3$~T and an initial distance of $d=4.7$~mm.}
    \label{fig:F_tan_over_delta_y}
\end{figure*}

The simulation procedure and all MMR parameters are identical to those of the stator tilt analysis in Section~\ref{ssec:manufacturing_errors_side_modes}. Figure~\ref{fig:F_tan_over_rotor_epsilon} shows the resulting equivalent perpendicular force for the tilted rotor of the Thread-MMR together with the tilted jewel of the Jewel-MMR, since the self-aligning rotor of the Jewel-MMR cannot exhibit a tilt defect of its own. Figure~\ref{fig:F_tan_over_delta_y} shows the corresponding results for a lateral displacement $\Delta y$ along the stator's magnetization direction: a displacement perpendicular to it results in no equivalent perpendicular force ($\widetilde{F}_\perp = 0$) due to the remaining symmetry of the configuration. As in the main part, subplot \textit{(a)} display multiple deflection angles $\varphi$ at the cup radius $R=\frac{5}{3}r$ for the Jewel-MMR and at the thread length $l=8r$ for the Thread-MMR, while subplots \textit{(b)} and \textit{(c)} explore multiple cup radii and thread lengths at a fixed deflection angle of $\varphi=30^\circ$. For the tilted rotor and the tilted jewel, the force curves closely resemble those of the tilted stator in Fig.~\ref{fig:F_tan_over_epsilon}, supporting the similarity argument of Section~\ref{ssec:types_of_manufacturing_errors}. The force curves for the lateral displacement are similar as well, however with an inverted dependence on the design parameters: the equivalent perpendicular force decreases with increasing cup radius and thread length and vanishes entirely in the limiting cases $R\to\infty$ and $l\to\infty$, since the flat jewel surface and the infinitely long thread restore the rotational symmetry of the configuration. Nevertheless, these limiting cases do not invalidate the design parameters chosen for the experiments: for the Thread-MMR, the dependence on the thread length is only weak and an arbitrarily long thread is not practically feasible, while for the Jewel-MMR, a flat jewel is technically possible but eliminates the geometric centering effect and thereby destabilizes the equilibrium position with respect to all tilt-type defects, as discussed in Section \ref{ssec:manufacturing_errors_side_modes}.

\begin{figure*}[pos=htbp]
    \centering
    \includegraphics[width=0.9\textwidth]{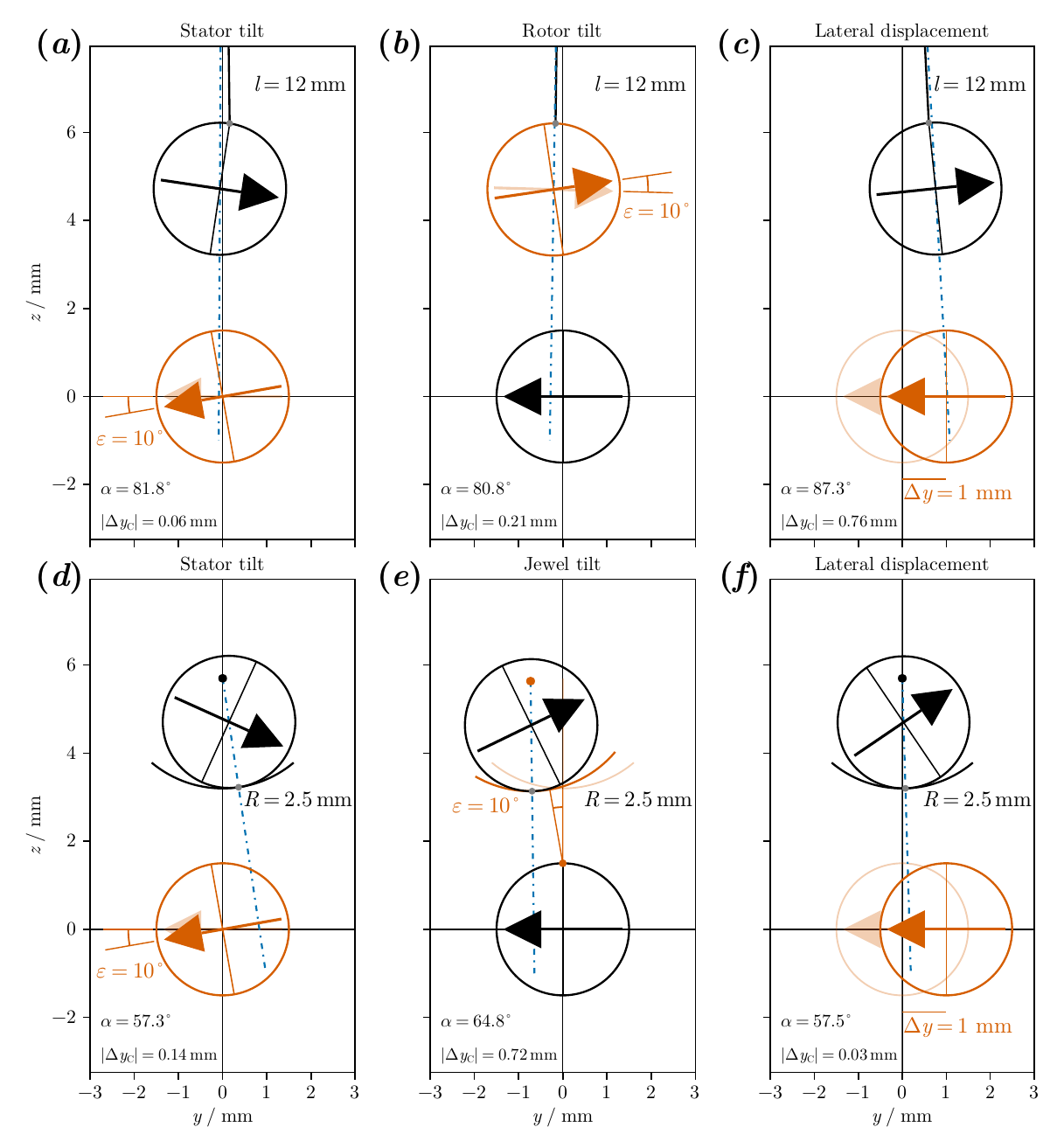}
    \vspace{-10pt}
    \caption{Simulated equilibrium positions of the rotor for all bearing error configurations of Fig.~\ref{fig:diagram-manufacturing-defects-schematic}, with a fixed error angle of $\varepsilon = 10^\circ$ for the tilt defects and a lateral displacement of $\Delta y = 1$~mm respectively. The light red shade indicates the initial state before the defect is applied, and the dash-dotted line marks the definition of the new axis of rotation. For each configuration, the absolute distance $\left\vert \Delta y_C \right\vert$ between the rotor's center and the $z$-axis as well as the angle $\alpha$ between the axis of rotation and the rotor's magnetization (ideally $\alpha = 90^\circ$) are given. The magnets of both MMRs have a radius of $r=1.5$~mm, density of $\rho=7800$~kg~m$^{-3}$, remanent magnetization of $B_\mathrm{r}=1.3$~T and an initial distance of $d=4.7$~mm.}
    \label{fig:equilibrium_positions_defects}
\end{figure*}

Finally, Fig.~\ref{fig:equilibrium_positions_defects} verifies the claim of Section \ref{ssec:types_of_manufacturing_errors} that every identified defect forces the rotor to oscillate about an axis that neither passes through both magnets' centers nor is perpendicular to the rotor's magnetization: for each error configuration of Fig.~\ref{fig:diagram-manufacturing-defects-schematic}, the computed equilibrium position for the rotor yields a center distance $\left\vert \Delta y_C \right\vert > 0$ from the $z$-axis and an angle $\alpha \neq 90^\circ$ between the axis of rotation and the rotor's magnetization.

\printcredits

\section*{Declaration of competing interest}
The authors declare that they have no known competing financial interests or
personal relationships that could have appeared to influence the work reported
in this paper.

\section*{Data availability}
Data will be made available on request.

\section*{Declaration of generative AI and AI-assisted technologies in the manuscript preparation process}
During the preparation of this work the authors used Claude Code by Anthropic in order to speed up the programming and writing process. After using this tool, the authors reviewed and edited the content as needed and take full responsibility for the content of the publication.

\printbibliography

\end{document}